\documentclass[pra,twocolumn,showpacs,preprintnumbers,amsmath,amssymb,superscriptaddress,
]{revtex4-2}

\usepackage{graphicx}
\usepackage{dcolumn}
\usepackage{bm}
\usepackage{color}
\usepackage{bbm}
\usepackage[colorlinks=true,citecolor=blue,linkcolor=blue,urlcolor=blue]{hyperref}
\usepackage{appendix}

\begin{document}

\title{Correlation Functions and Photon-Photon Interactions Controlled by a Giant Atom
}

\author{Yanjin Yue}
\affiliation{School of Physics, Sun Yat-sen University, Guangzhou 510275, China}
\author{Rui-Yang Gong}
\affiliation{School of Physics, Sun Yat-sen University, Guangzhou 510275, China}
\author{Shengyong Li}
\affiliation{Department of Automation, Tsinghua University, Beijing 100084, China}
\author{Ze-Liang Xiang}
\email{xiangzliang@mail.sysu.edu.cn}
\affiliation{School of Physics, Sun Yat-sen University, Guangzhou 510275, China}
\affiliation{State Key Laboratory of Optoelectronic Materials and Technologies, Sun Yat-sen University, Guangzhou 510275, China}
\date{\today}

\begin{abstract}
    Waveguide quantum electrodynamics (WQED) provides a powerful platform for exploring quantum optical phenomena by enhancing atom-photon interactions through photon confinement in a waveguide. Here we investigate the photon-scattering dynamics of a weak coherent pulse incident from the left on a giant atom coupled to a bidirectional waveguide, focusing on effects absent in the small-atom approximation. Using an extended input-output formalism, we calculate the relevant correlation functions and show that the competition between two scattering processes is governed by the ratio of the pulse width to the atomic lifetime, leading to time-dependent switching between bunching and antibunching. In addition, tuning the phase accumulated between the two coupling points of the giant atom allows the photon statistics to be switched among three distinct regimes, each with a finite phase bandwidth. We also discuss the experimental feasibility in superconducting circuits. Our results provide a route toward giant-atom-based control of photon pulses and potential applications in quantum control.
\end{abstract}

\maketitle

\section{Introduction}\label{sec:section1}

Waveguide quantum electrodynamics (WQED)~\cite{RevModPhys.95.015002,RevModPhys.89.021001} studies systems composed of natural or artificial atoms coupled to a variety of optical and microwave waveguides. By confining photons within the waveguide, WQED enables much stronger atom-photon interactions than in free space~\cite{RevModPhys.95.015002}, therefore facilitating applications such as photon manipulation~\cite{lund2024subtraction,rosenblum2016extraction,PhysRevA.97.062318} and quantum networks~\cite{PhysRevLett.117.240501,doi:10.1126/sciadv.abb8780,lin2022deterministic,PhysRevApplied.17.054021,PhysRevX.7.011035}. Under the dipole approximation~\cite{Scully_Zubairy_1997,agarwal2012quantum}, the atom can be treated as a pointlike emitter, commonly referred to as a small atom. The properties of the small atom coupled to a waveguide have been widely studied, including strongly correlated photons~\cite{tian2025disorder,PhysRevLett.98.153003,PhysRevLett.121.143601}, topological effects~\cite{PhysRevA.111.033716,poshakinskiy2021quantum,PhysRevLett.128.203602,ylrq-98wl}, and nonlinear photon-photon interactions~\cite{le2021experimental,vcepulkovskis2017nonlinear}. Some of these phenomena have recently been observed in experiments~\cite{le2022dynamical,prasad2020correlating}. 

In contrast to the simple single-point coupling of a small atom, an atom can also couple to a waveguide at two or more spatially separated points. In this case, the dipole approximation breaks down, giving rise to richer and more intriguing physical phenomena. Such a system is referred to as a giant atom~\cite{kockum2021quantum,andersson2019non,PhysRevA.90.013837,frisk2017quantum}. Giant atoms have been experimentally realized in several systems, including magnons~\cite{wang2022giant} and superconducting qubits coupled to transmission lines~\cite{kannan2020waveguide,PhysRevX.13.021039} or surface acoustic waves~\cite{doi:10.1126/science.1257219,andersson2019non}. In terms of applications, giant atoms can support tunable chiral bound states~\cite{PhysRevLett.126.043602} and serve as quantum gates~\cite{chen2026scalable}, quantum routers~\cite{Wang:21,PhysRevA.111.043713,gong2024tunable}, and quantum batteries~\cite{yan2025giant}. In particular, the phase accumulated between different coupling points can lead to nonreciprocal and chiral photon scattering~\cite{gong2024tunable,chen2022nonreciprocal} as well as decoherence-free interactions~\cite{PhysRevLett.120.140404,PhysRevA.107.023705,PhysRevResearch.2.043184,PhysRevA.105.023712}, whereas the associated time delay gives rise to non-Markovian effects~\cite{PhysRevLett.133.063603,qiu2023collective}. Some of these properties are influenced by the characteristics of the incident photons and can already emerge in the few-photon input regime. Modeling incident photons as wavepackets is more general than treating them as monochromatic, i.e., taking the spectral amplitude to be a delta function, and it allows one to access the temporal dynamics of the photons. This perspective has already been explored both theoretically and experimentally in the small-atom case~\cite{le2021experimental,le2022dynamical}. These developments naturally motivate the question of whether new physical phenomena and potential applications can emerge when the few-photon pulse with a spectral amplitude is mediated by a giant atom.


To characterize such phenomena, correlation functions provide an essential tool, as they directly probe the intensity and statistical properties of photons~\cite{Scully_Zubairy_1997}. In WQED and related systems such as cavity quantum electrodynamics systems~\cite{walther2006cavity,RevModPhys.73.565} and circuit quantum electrodynamics systems~\cite{RevModPhys.93.025005,GU20171,haroche2020cavity}, correlation functions are closely connected to nonclassical effects, including the structures of two-photon and multi-photon states~\cite{mahmoodian2020dynamics,gu2023correlated,liang2018observation}, and can serve as indicators of photon blockade~\cite{PhysRevLett.121.043602,PhysRevLett.121.043601,bin2018two,lu2025chiral,PhysRevLett.122.243602,huang2022exceptional,7frd-pf1m}. A variety of theoretical approaches have been developed to evaluate correlation functions in WQED systems, such as the quantum trajectory method~\cite{PhysRevA.106.013714,PhysRevA.93.062104,daley2014quantum},
the Green’s function method~\cite{PhysRevA.93.013828,PhysRevA.108.053703,ke2019inelastic,PhysRevA.84.063803}, and the Schrödinger equation method~\cite{PhysRevA.85.015803,nysteen2015scattering,PhysRevA.96.053805}. In the Schrödinger-equation approach, one solves for the photon wavefunctions and matches the solutions in different regions through boundary conditions, from which the correlation functions are then obtained. An alternative is the extended input-output formalism, which combines scattering theory and input-output theory. The approach evaluates correlation functions through Langevin equations without explicitly solving for the full wavefunctions~\cite{fan2010input}, and it has been experimentally validated for a single small atom~\cite{le2022dynamical}. However, it cannot be straightforwardly extended to the giant-atom case, because the original treatment starts from the simpler chiral-waveguide solution and obtains the bidirectional result through minimal substitutions only—an approach that breaks down for giant atoms. 

In this work, we study a WQED system in which a weak coherent pulse (with mean photon number $\ll 1$) of finite temporal width is incident from the left, i.e., right-propagating direction, on a bidirectional waveguide coupled to a two-level giant atom. To evaluate the correlation functions, we extend the combined method of scattering theory and the input-output formalism by treating the bidirectional-waveguide case directly, thereby making the method applicable to giant atoms. For the transmitted field, the two-time correlation functions reveal that the competition between two processes causes switching between bunching and antibunching as a function of absolute time. This competition is controlled by the ratio of the pulse width to the atomic lifetime, which we hereafter simply refer to as the ratio. The first process involves the interaction of a single photon with the atom, whereas the second involves two photons forming a bound state and being emitted simultaneously. Another effect arises from the phase accumulated between two coupling points, which provides an additional tunable degree of freedom. At appropriately chosen times, tuning this phase switches the output field among bunching, antibunching, and coherent regimes, each of which persists over a finite phase bandwidth. This phenomenon is absent for small atoms and may have potential experimental applications. Finally, we discuss the feasibility of observing these effects on a superconducting platform, including possible measurement protocols and realistic parameter ranges. 

This paper is organized as follows. In Sec.~\ref{sec:Model}, the model and corresponding Hamiltonian are introduced. Section~\ref{sec:Derivations} derives the single-photon and two-photon scattering matrices, and then the second-order correlation functions. By analyzing these correlation functions, we illustrate the competition between the two processes in Sec.~\ref{sec:Temporal Photon Dynamics}, and show that by tuning the phase, the field can be switched among three distinct regimes in Sec.~\ref{sec:Phase Control}. Section~\ref{sec:Discussion} discusses the feasibility of implementing our proposal in experiments. Finally, Sec.~\ref{sec:Conclusion} offers conclusions of our work. 


\section{\bf Model}\label{sec:Model}

\begin{figure}
  \centering
  \includegraphics[width=1\linewidth]{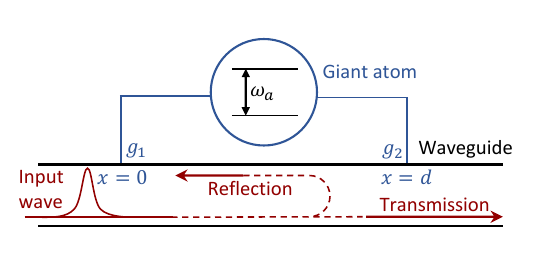} \caption{Schematic of a two-level giant atom coupled to a waveguide at two points with coupling strengths $g_1$ and $g_2$. A weak coherent pulse is incident from the left.}
  \label{fig:illustrator}
\end{figure}

We consider a giant atom coupled to a bidirectional waveguide, as shown in Fig.~\ref{fig:illustrator}, where a weak coherent pulse is incident in the right-propagating direction. The Hamiltonian of the giant-atom-waveguide system can be written as 
\begin{equation}
\label{eq:Hamiltonian}
\begin{aligned}
H = &\frac{1}{2}(\omega_a-\omega_0)\sigma_z
+ \int_{-\infty}^{\infty}\mathrm{d}\omega \cdot (\omega-\omega_0)
\left(r_\omega^\dagger r_\omega + \ell_\omega^\dagger \ell_\omega\right) \\
& + \int_{-\infty}^{\infty}\mathrm{d}\omega \Big\{
\left[g_1(t)+g_2(t)e^{ik_0 d}e^{i(\omega-\omega_0)\frac{d}{v_g}}\right]\sigma_{+}r_\omega \\
&
+\left[g_1(t)+g_2(t)e^{-ik_0 d}e^{-i(\omega-\omega_0)\frac{d}{v_g}}\right]\sigma_{+}\ell_\omega
+\mathrm{h.c.}\Big\},
\end{aligned}
\end{equation}
where $\omega_a$ is the atomic frequency, $\sigma_z$ is the Pauli-z operator, and $\sigma_+$ is the Pauli raising operator. Throughout this paper, we set $\hbar=1$. The operators $r_\omega $ and $\ell_\omega$ are bosonic annihilation operators that represent right-propagating and left-propagating photons, respectively. The two coupling points are located at $x=0$ and $x=d$, with complex coupling strengths $g_1$ and $g_2$, respectively. 

Because the incident photons are concentrated around $\omega_0$, we adopt the linear-waveguide approximation~\cite{RevModPhys.89.021001} and extend the frequency integral to infinity. Since the total number of excitations number
\begin{equation}
N_{\Sigma}=\frac{1}{2}\sigma_z
+ \int_{-\infty}^{\infty}\mathrm{d}\omega 
\left(r_\omega^\dagger r_\omega + \ell_\omega^\dagger \ell_\omega\right)
\end{equation}
is conserved, i.e., $[H,N_{\Sigma}]=0$, we shift the Hamiltonian by $\omega_0$~\cite{fan2010input}. 

In Eq.~\eqref{eq:Hamiltonian}, the first term describes the atom, the second term corresponds to the bidirectional waveguide, and the third term represents the interaction between the giant atom and the waveguide field. Here $k_0$ is the wave number at frequency $\omega_0$, and $v_g$ is the corresponding group velocity. 


\section{\bf Methods}\label{sec:Derivations}

For this system, we first derive the single- and two-photon scattering matrices and then evaluate the second-order correlation functions under a weak coherent pulse for arbitrary directions.

\subsection{\label{sec:Single-photon Scattering Matrices}Single-photon Scattering Matrices}

The single-photon scattering matrices for a photon incident from the right-propagating mode are defined as
\begin{equation}
\label{eq:single photon S}
    \langle0|\ell_{out}(p)r_{in}^{\dagger}(k)|0\rangle,
    \qquad
    \langle0|r_{out}(p)r_{in}^{\dagger}(k)|0\rangle.
\end{equation}
The former describes reflection, while the latter describes transmission. Here, $|0\rangle$ denotes the joint vacuum state of the waveguide fields (including both right- and left-propagating modes) and the atom, i.e., no photons are present and the atom is in its ground state. 

In Eq.~\eqref{eq:single photon S}, $r_{in}^{\dagger}(k)$, $\ell_{out}(p)$ and $r_{out}(p)$ are scattering operators, where $k$ and $p$ represent the corresponding photon frequency. They are related to the input-output operators by~\cite{fan2010input}
\begin{align}
\label{eq:tk-relation}
    r_{\mathrm{in}}(t)&=\frac{1}{\sqrt{2\pi}}\int \mathrm{d}k\,r_{\mathrm{in}}(k)e^{-ikt},\\
    r_{\mathrm{out}}(t)&=\frac{1}{\sqrt{2\pi}}\int \mathrm{d}k\,r_{\mathrm{out}}(k)e^{-ikt},
\end{align}
and analogous relations hold for $\ell$. The input-output relations for the right- and left-propagating fields read
\begin{equation}
\label{eq:rl-input-output}
\begin{aligned}
r_{\mathrm{out}}(t)
&= r_{\mathrm{in}}(t)
  - i\sqrt{2\pi}g_1^{*}\sigma_{-}(t)
  - i\sqrt{2\pi}g_2^{*}\sigma_{-}(t)e^{-i\varphi_0},
\\
\ell_{\mathrm{out}}(t)
&= \ell_{\mathrm{in}}(t)
  - i\sqrt{2\pi}g_1^{*}\sigma_{-}(t)
  - i\sqrt{2\pi}g_2^{*}\sigma_{-}(t)e^{i\varphi_0},
\end{aligned}
\end{equation}
where $\varphi_0=k_0d$. Here, we neglect the time delay $\tau=d/{v_g}$ between the two coupling points on the operators, while retaining the corresponding phase accumulation. The derivation of Eq.~\eqref{eq:rl-input-output} and a brief discussion of this approximation are given in Appendix~\ref{sec:Input-output Relation}.

Using Eqs.~\eqref{eq:tk-relation} and~\eqref{eq:rl-input-output}, the transmission scattering matrix can be written as 
\begin{equation}
\label{eq:transmission solving}
\begin{aligned}
&\langle0|r_{\mathrm{out}}(p)r_{\mathrm{in}}^{\dagger}(k)|0\rangle\\
=&\,\frac{1}{\sqrt{2\pi}}\int dt e^{ipt}[\langle0|r_{\mathrm{in}}(t)r_{\mathrm{in}}^{\dagger}(k)|0\rangle\\
&\,- i\sqrt{2\pi}(g_1^{*}+g_2^{*}e^{-i\varphi_0})\langle0|\sigma_{-}(t)r_{\mathrm{in}}^{\dagger}(k)|0\rangle].
\end{aligned}
\end{equation}

We therefore need $\langle0|r_{\mathrm{in}}(t) r_{\mathrm{in}}^{\dagger}(k)|0\rangle$ and $\langle0|\sigma_{-}(t) r_{\mathrm{in}}^{\dagger}(k)|0\rangle$. The former follows directly from Eq.~\eqref{eq:tk-relation}. The latter is obtained from the Langevin equation for $\sigma_-$, or $\langle0|\sigma_{-}(t) r_{\mathrm{in}}^{\dagger}(k)|0\rangle$, specifically. Denoting $X(t)=\langle0|\sigma_-(t)r_{\mathrm{in}}^{\dagger}(k)|0\rangle$, the equation of $X(t)$ is given by   
\begin{equation}
\label{eq:X equation}
\begin{aligned}
\dot{X}
=& -{i}\Delta X
-i(g_1+g_2e^{i\varphi_{0}}) e^{-ikt}\\
&-2\pi \big(|g_1|^2+|g_2|^2\big)X
-4\pi \mathrm{Re}(g_1 g_2^{*})e^{i\phi_{0}}X,
\end{aligned}
\end{equation}
where $\Delta=\omega_a-\omega_0$ is the detuning between the photon and the atom. The derivation of Eq.~\eqref{eq:X equation} is provided in Appendix~\ref{sec:Derivation of the Langevin Equation}. Its solution is 
\begin{equation}
\label{eq:sk time}
X(t)=\frac{(g_1+g_2e^{i\phi_0})e^{-ikt}}{k-\Delta+2\pi i(|g_1|^2+|g_2|^2)+4\pi i\mathrm{Re}(g_1 g_2^{*})e^{i\phi_{0}}}.
\end{equation}
Taking the Fourier transform of $X(t)$, we have  
\begin{equation}
\label{eq:sr delta}
\langle0|\sigma_-(p)r_{\mathrm{in}}^{\dagger}(k)|0\rangle=\delta(p-k)s_r(k),
\end{equation}
with
\begin{equation}
\label{eq:sr}
s_r(k)=\frac{\sqrt{2\pi}(g_1+g_2e^{i\phi_0})}{k-\Delta+2\pi i(|g_1|^2+|g_2|^2)+4\pi i\mathrm{Re}(g_1 g_2^{*})e^{i\phi_{0}}},
\end{equation}
which is the excitation amplitude for a single photon incident from the right-propagating mode and absorbed by the atom. 

Substituting Eq.~\eqref{eq:tk-relation} and Eq.~\eqref{eq:sk time} into Eq.~\eqref{eq:transmission solving}, we obtain
\begin{equation}
\label{eq:r-transmission-delta}
\langle0|r_{\mathrm{out}}(p)r_{\mathrm{in}}^{\dagger}(k)|0\rangle=\delta(k-p)t_r,
\end{equation}
where
\begin{equation}
\label{eq:r-transmission}
t_r=\frac{k-\Delta-4\pi  g_1g^*_2\sin(\phi_0)}{k-\Delta+2\pi i(|g_1|^2+|g_2|^2)+4\pi i\mathrm{Re}(g_1 g_2^{*})e^{i\phi_{0}}}
\end{equation}
is the transmission amplitude. The delta function enforces frequency conservation, and the subscript $r$ indicates that the incident photon is right propagating. This result agrees with the transmission amplitude in Ref.~\cite{chen2022nonreciprocal}.

Similarly, 
\begin{equation}
\label{eq:r-reflection-delta}
\langle0|l_{\mathrm{out}}(p)r_{\mathrm{in}}^{\dagger}(k)|0\rangle=\delta(k-p)r_r,
\end{equation}
where
\begin{equation}
\label{eq:r-reflection}
r_r=\frac{-2\pi i[|g_1|^2+|g_2|^2e^{2i\phi_0}+2\mathrm{Re}(g_1 g_2^{*})e^{i\phi_0}] }{k-\Delta+2\pi i(|g_1|^2+|g_2|^2)+4\pi i\mathrm{Re}(g_1 g_2^{*})e^{i\phi_{0}}}
\end{equation}
is the reflection amplitude. One can verify that $|r_r|^2+|t_r|^2=1$, as no dissipation channel other than the waveguide is included.

The transmission and reflection amplitudes for a photon incident from the left-propagating mode are denoted as $t_{\ell}$ and $r_{\ell}$. We list the results in Appendix~\ref{sec:Single-photon Scattering Matrices for Left-propagating Mode}, as they will be useful in the following subsection.


\subsection{Two-photon Scattering Matrices}
\label{sec:Two-photon Scattering Matrices}

In this section, we derive the two-photon scattering matrix
\begin{equation}
\label{eq:TSM-rr}
\langle0|r_{\mathrm{out}}(p_1)r_{\mathrm{out}}(p_2)r_{\mathrm{in}}^{\dagger}(k_1)r_{\mathrm{in}}^{\dagger}(k_2)|0\rangle,
\end{equation}
which describes the process in which both incident photons are transmitted. The corresponding results for the other two cases, namely, both photons reflected and one photon transmitted while the other one is reflected, are presented in Appendix~\ref{sec:Derivation of Other Two-photon Scattering Matrices}. 

We begin by inserting the normalization condition
\begin{equation}
\label{eq:normalization condition}
\int_{-\infty}^{\infty} \! dk\;\Big[r_{\mathrm{in}}^{\dagger}(k)|0\rangle\langle 0|r_{\mathrm{in}}(k)+\ell_{\mathrm{in}}^{\dagger}(k)|0\rangle\langle 0|\ell_{\mathrm{in}}(k)\Big]= 1,
\end{equation}
and simplify the single-photon scattering matrices using Eqs.~\eqref{eq:r-transmission-delta}-\eqref{eq:r-reflection}. This gives 
\begin{equation}
\begin{aligned}
&\langle0|r_{\mathrm{out}}(p_1)r_{\mathrm{out}}(p_2)r_{\mathrm{in}}^{\dagger}(k_1)r_{\mathrm{in}}^{\dagger}(k_2)|0\rangle
=\\&\;\Big[t_r(p_1)\langle 0|r_{\mathrm{in}}(p_1)r_{\mathrm{out}}(p_2)r_{\mathrm{in}}^{\dagger}(k_1)r_{\mathrm{in}}^{\dagger}(k_2)|0\rangle
\\&+r_{\ell}(p_1)\langle 0|\ell_{\mathrm{in}}(p_1)r_{\mathrm{out}}(p_2)r_{\mathrm{in}}^{\dagger}(k_1)r_{\mathrm{in}}^{\dagger}(k_2)|0\rangle\Big].
\end{aligned}
\end{equation}

Using Eq.~(\ref{eq:tk-relation}) together with the input-output relation Eq.~(\ref{eq:rl-input-output}), we further obtain
\begin{equation}
\label{eq:rrrr}
\begin{aligned}
&\langle 0|r_{\mathrm{in}}(p_1)r_{\mathrm{out}}(p_2)r_{\mathrm{in}}^{\dagger}(k_1)r_{\mathrm{in}}^{\dagger}(k_2)|0\rangle
=\frac{1}{\sqrt{2\pi}}\int dt e^{ip_2t}\\
&\Big[ t_r(p_1)\big\langle 0\big| r_{\mathrm{in}}(p_1)r_{\mathrm{in}}(t)r_{\mathrm{in}}^{\dagger}(k_1)r_{\mathrm{in}}^{\dagger}(k_2)\big|0\big\rangle-t_r(p_1)\\
&\times i\sqrt{2\pi}\big(g_2^{*}e^{-i\phi_0}+g_1^{*}\big)\big\langle 0\big| r_{\mathrm{in}}(p_1)\sigma_{-}(t)r_{\mathrm{in}}^{\dagger}(k_1)r_{\mathrm{in}}^{\dagger}(k_2)\big|0\big\rangle\\
&+ r_{\ell}(p_1)\big\langle 0\big| \ell_{\mathrm{in}}(p_1)r_{\mathrm{in}}(t)r_{\mathrm{in}}^{\dagger}(k_1)r_{\mathrm{in}}^{\dagger}(k_2)\big|0\big\rangle- r_{\ell}(p_1)\\
&\times i\sqrt{2\pi}\big(g_2^{*}e^{-i\phi_0}+g_1^{*}\big)\big\langle 0\big| \ell_{\mathrm{in}}(p_1)\sigma_{-}(t)r_{\mathrm{in}}^{\dagger}(k_1)r_{\mathrm{in}}^{\dagger}(k_2)\big|0\big\rangle \Big].
\end{aligned}
\end{equation}

The first term is obtained directly from the commutation relations and Eq.~\eqref{eq:tk-relation},
\begin{equation}
\begin{aligned}
&\big\langle 0\big| r_{\mathrm{in}}(p_1)r_{\mathrm{in}}(t)r_{\mathrm{in}}^{\dagger}(k_1)r_{\mathrm{in}}^{\dagger}(k_2)\big|0\big\rangle=
\frac{1}{\sqrt{2\pi}}
\\&\times[e^{-ik_1t}\delta(p_1-k_2)\delta(p_2-k_1)
\\&+e^{-ik_2t}\delta(p_2-k_2)\delta(p_1-k_1)].
\end{aligned}
\end{equation}
The third term vanishes because of Eq.~\eqref{eq:lr product}. We are therefore left with the second and fourth terms, which can be obtained from the corresponding Langevin equations. The results are
\begin{equation}
\label{eq:rsrr}
\begin{aligned}
&\big\langle 0 \big| r_{\mathrm{in}}(p_{1}) \sigma_{-}(p_2)
r_{\mathrm{in}}^{\dagger}(k_{1}) r_{\mathrm{in}}^{\dagger}(k_{2}) \big| 0 \big\rangle\\
=&-\frac{1}{\pi}\delta(k_{1}+k_{2}-p_{1}-p_{2})s_r({p_2})s_r^*({p_1}) [s_r({k_1})+s_r({k_2})]
\\&+ s_r({k_1})\delta(k_{2}-p_{1})\delta(k_{1}-p_{2})
\\&+s_r({k_2})\delta(k_{1}-p_{1})\delta(k_{2}-p_{2}),
\end{aligned}
\end{equation}
and
\begin{equation}
\label{eq:lsrr}
\begin{aligned}
&\big\langle 0 \big| \ell_{\mathrm{in}}(p_{1}) \sigma_{-}(p_2)
r_{\mathrm{in}}^{\dagger}(k_{1}) r_{\mathrm{in}}^{\dagger}(k_{2}) \big| 0 \big\rangle\\
=&-\frac{1}{\pi}\delta(k_{1}+k_{2}-p_{1}-p_{2})s_r({p_{2}})s^{*}_{\ell}({p_{1}})[s_r({k_1})+s_r({k_2})].
\end{aligned}
\end{equation}
The derivation is presented in Appendix~\ref{sec:Both Photons are Transmitted}.

Equation~\eqref{eq:rsrr} describes the scattering process for two photons incident from the right-propagating mode with frequencies $k_1$ and $k_2$. One photon is absorbed by the atom with frequency $p_2$, while the other is transmitted with frequency $p_1$. The first term in Eq.~\eqref{eq:rsrr} represents an effective photon–photon interaction mediated by the atom, whereas the other two terms correspond to the case where no such interaction occurs.

Equation~\eqref{eq:lsrr} describes the corresponding process in which the outgoing photon with frequency $p_1$ appears in the left-propagating mode. In contrast to Eq.~\eqref{eq:rsrr}, Eq.~\eqref{eq:lsrr} contains only the atom-mediated photon-photon interaction term, because reflection of a right-incident photon can occur only through interaction with the atom. 

Substituting Eqs.~\eqref{eq:rsrr} and~\eqref{eq:lsrr} into Eq.~\eqref{eq:rrrr}, we arrive at
\begin{equation}
\label{eq:rrrr-result}
\begin{aligned}
&\langle0|r_{\mathrm{out}}(p_1)r_{\mathrm{out}}(p_2)r_{\mathrm{in}}^{\dagger}(k_1)r_{\mathrm{in}}^{\dagger}(k_2)|0\rangle
\\
=&t_r({k_2})
t_r({k_{1}})\delta(k_{2}-p_{1})\delta(k_{1}-p_{2})
\\&+t_r({k_{1}})t_r({k_{2}})\delta(k_{1}-p_{1})\delta(k_{2}-p_{2})
\\&+i\sqrt{2\pi}\big(g_2^{*}e^{-i\phi_0}+g_1^{*}\big)\frac{1}{\pi}\delta(k_{1}+k_{2}-p_{1}-p_{2})\\&\times s_r({p_2})s_r({p_1})[s_r({k_1})+s_r({k_2})]\frac{g^*_1+g^*_2e^{-i\phi_0}}{g_1+g_2e^{i\phi_0}}.
\end{aligned}
\end{equation}
This is the desired two-photo scattering matrix for the transmission channel (also see Appendix~\ref{sec:Both Photons are Transmitted}). As in Eq.~\eqref{eq:rsrr}, the first two terms describe independent single-photon transmission processes and arise from exchange symmetry, while the last term represents the effective photon–photon interaction mediated by the atom, as indicated by the factor $s_r$. This contribution conserves the total frequency and depends on the overall dissipative coupling strength $g_2^{*}e^{i\phi_0}+g_1^{*}$. 

When $\varphi_0=0$, the two-photon scattering matrix for the giant atom reduces to that of a two-level small atom. Moreover, if $g_1+g_2$ is taken to be real, our result becomes identical to that reported in Ref.~\cite{fan2010input}, with the correspondence $\frac{1}{\tau^{'}}=\pi|g_1+g_2|^2$.


\subsection{Second-order Correlation Functions under Coherent Driving }
\label{sec:Second-order Correlation Functions under Coherent Driving}

In this section, we relate the two-photon scattering matrix to the second-order correlation function under weak coherent driving. The input field is a weak coherent pulse incident from the right-propagating mode, with spectral amplitude $f(k)$. Its temporal profile is obtained by Fourier transformation~\cite{PhysRevX.5.041017}. For a two-level small atom, the corresponding derivation was given in Ref.~\cite{le2021experimental}. Here we generalize the main steps and results to the case of the giant atom, and show how the correlation function can be normalized in a well-defined way. 

We begin with the Fock-state expansion of the weak coherent input state $|\Psi_{\mathrm{in}}^{(\alpha)}\rangle$
\begin{equation}
|\Psi_{\mathrm{in}}^{(\alpha)}\rangle=|0\rangle+\alpha|\Psi_{\mathrm{in}}^{(1)}\rangle+\frac{\alpha^{2}}{\sqrt{2}}|\Psi_{\mathrm{in}}^{(2)}\rangle+\mathcal{O}(\alpha^{3}),
\end{equation}
where the superscript $\alpha$ indicates the coherent state, and $(1)$ and $(2)$ label the one- and two-photon Fock sectors, respectively. Since the coherent drive is weak ($\alpha\approx0$) and the second-order correlation function involves up to the two-photon process, terms of order $\alpha^3$ and higher can be neglected. 

The single- and two-photon input states are
\begin{equation}
\begin{aligned}
|\Psi_{\mathrm{in}}^{(1)}\rangle&=\int d\omega f(k)r_{\mathrm{in}}^{\dagger}(k)|0\rangle,\\
|\Psi_{\mathrm{in}}^{(2)}\rangle&=\int\int\frac{dk_{1}dk_{2}}{\sqrt{2}}f(k_{1})f(k_{2})r_{\mathrm{in}}^{\dagger}(k_{1})r_{\mathrm{in}}^{\dagger}(k_{2})|0\rangle,
\end{aligned}\end{equation}
where $r_{\mathrm{in}}(k)$ is defined in Eq.~\eqref{eq:tk-relation}.
 
The corresponding output states can be written as~\cite{le2021experimental}
\begin{equation}
\label{eq:output state}
\begin{aligned}
|\Psi_{\mathrm{out}}^{(1)}\rangle
&=\sum_{\mu=t,r}\int\int dpdk  f(k) S_{pk}^{\mu} a_{\mathrm{in}}^{\mu\dagger}(p)|0\rangle,\\
|\Psi_{\mathrm{out}}^{(2)}\rangle
&=\sum_{\mu\mu^{\prime}=t,r}\int\int \frac{dp_{1}dp_{2}}{2\sqrt{2}}
\int\int dk_{1}dk_{2} f(k_{1})f(k_{2}) \\
&\quad \times S_{p_{1}p_{2}k_{1}k_{2}}^{\mu\mu^{\prime}}
a_{\mathrm{in}}^{\mu\dagger}(p_{1})a_{\mathrm{in}}^{\mu^{\prime}\dagger}(p_{2})|0\rangle.
\end{aligned}
\end{equation}
Here $\mu$ labels the output direction, with $t$ and $r$ denoting transmission and reflection, respectively. For notational convenience, we define
\begin{equation}
a_{\mathrm{in}}^{t}(p)= r_{\mathrm{in}}(p),\qquad a_{\mathrm{in}}^{r}(p)= \ell_{\mathrm{in}}(p).
\end{equation}
The relevant scattering matrices are 
\begin{equation}
\label{eq:scattering matrix correspondance}
\begin{aligned}
 S_{pk}^{r}&=\langle0|\ell_{out}(p)r_{in}^{\dagger}(k)|0\rangle=\chi^t(k)\delta(p-k),
\qquad\\
 S_{pk}^{t}&=\langle0|r_{out}(p)r_{in}^{\dagger}(k)|0\rangle=\chi^r(k)\delta(p-k),
\\
S_{p_{1}p_{2}k_{1}k_{2}}^{tt}&=\langle0|r_{\mathrm{out}}(p_1)r_{\mathrm{out}}(p_2)r_{\mathrm{in}}^{\dagger}(k_1)r_{\mathrm{in}}^{\dagger}(k_2)|0\rangle,
\end{aligned}
\end{equation}
where $\chi^t(k)=t_r(k)$ and $\chi^{r}(k)=r_r(k)$. 

Projecting Eq.~\eqref{eq:output state} onto the desired output directions and times gives the single- and two-photon wavefunctions~\cite{le2021experimental},
\begin{equation}
\label{eq:output result1}
\begin{aligned}
\psi_{\mu_1}^{(1)}(t)&=\frac{1}{\sqrt{2\pi}}  \int\int dp dk e^{-ip t}f(k)S_{pk}^{\mu_1},
\\
\psi_{\mu_2\mu^{\prime}_2}^{(2)}(t,t+\tau)&=\frac{1}{2\pi}\int\int\frac{dp_{1}dp_{2}}{\sqrt{2}}\int\int dk_{1}dk_{2}
\\&\quad\{e^{-i[p_1t+p_2(t+\tau)]}f(k_{1})f(k_{2})   S_{p_{1}p_{2}k_{1}k_{2}}^{\mu_2\mu^{\prime}_2}\}.
\end{aligned}\end{equation}

Substituting the scattering matrices, Eqs.~\eqref{eq:scattering matrix correspondance} and~\eqref{eq:llrr-result}-\eqref{eq:rlrr-result}, into Eq.~\eqref{eq:output result1}, we obtain 
\begin{equation}
\begin{aligned}
\psi_{\mu_1}^{(1)}(t)=&\frac{1}{\sqrt{2\pi}}\int dk e^{-ikt}f(k)\chi^{\mu_1}(k),\\
\psi_{\mu_2\mu_2^{\prime}}^{(2)}(t,t+\tau)=&\frac{1}{\sqrt{2}\pi}\int dk_{1}f(k_{1})e^{-ik_{1}t}\chi^{\mu_2}(k_{1}) \\
&\times\int dk_{2}f(k_{2})e^{-ik_{2}(t+\tau)}\chi^{\mu_2^{\prime}}(k_{2})\\
& + N_{\mu_2\mu_2^{\prime}}(t,t+\tau),
\end{aligned}
\end{equation}
where
\begin{equation}
\label{eq:two-photon nonlinearity}
\begin{aligned}
N_{\mu_2\mu_2^{\prime}}(t,t+\tau)=&-\frac{iDe^{iC|\tau|}B_{c,\mu_2\mu_2^{\prime}} }{\sqrt{2}}\\
&\times\{\int dk f(k)e^{-ik(t+\frac{\tau}{2}-\frac{|\tau|}{2})}r_r(k)\}^2.
\end{aligned}
\end{equation}
This term represents the two-photon nonlinearity generated by the atom-mediated photon–photon interaction. The constants are given in Eq.~\eqref{eq:constants}, and the derivation is presented in Appendix~\ref{sec:Derivation of the Two-photon Nonlinearity}. 

Because $\mathrm{Re}(iC)<0$, the nonlinear contribution is appreciable only near $\tau = 0$. For $\mathrm{Re}(iC)|\tau|\gg 0$, it becomes negligible, and the two-photon wavefunction factorizes into a product of two single-photon wavefunctions, 
\begin{equation}
\label{eq:2nd to 1st}
\psi_{\mu_2\mu_2^{\prime}}^{(2)}(t,t+\tau)\propto\psi_{\mu_2}^{(1)}(t)\psi_{\mu'_2}^{(1)}(t+\tau).
\end{equation}
Its squared modulus then gives the normalization factor of the second-order correlation function, namely, the product of the single-photon intensity spectrum
\begin{equation}
I_{\mu_2}(t)I_{\mu_2^{\prime}}(t+\tau)=|\psi_{\mu_2}^{(1)}(t)|^2|\psi_{\mu'_2}^{(1)}(t+\tau)|^2.
\end{equation}

The unnormalized second-order correlation function is
\begin{equation}
\label{eq:2nd correlation}
G_{\mu_2\mu_2^{\prime}}^{(2)}(t,t+\tau)=|\psi_{\mu_2\mu_2^{\prime}}^{(2)}(t,t+\tau)|^2+\mathcal{O}(\alpha^{2}).
\end{equation}
In this work, we use the normalized correlation function~\cite{kubanek2008two}
\begin{equation}
\label{eq:2nd correlation normalized}
C_{\mu_2\mu_2^{\prime}}^{(2)}(t,t+\tau)=G_{\mu_2\mu_2^{\prime}}^{(2)}(t,t+\tau)-I_{\mu_2}(t)I_{\mu_2^{\prime}}(t+\tau),
\end{equation}
rather than the more common ratio
\begin{equation}
G_{\mu_2\mu_2^{\prime}}^{(2)}(t,t+\tau)/I_{\mu_2}(t)I_{\mu_2^{\prime}}(t+\tau),
\end{equation}
in order to avoid divergences when $I_{\mu_2}(t)I_{\mu_2^{\prime}}(t+\tau)=0$ but $G_{\mu_2\mu_2^{\prime}}^{(2)}(t,t+\tau)\neq0$~\cite{gulfam2018highly}. With this convention, $C_{\mu_2\mu_2^{\prime}}^{(2)}(t,t+\tau)>0$ indicates bunching, $C_{\mu_2\mu_2^{\prime}}^{(2)}(t,t+\tau)=0$ corresponds to coherent output, and $C_{\mu_2\mu_2^{\prime}}^{(2)}(t,t+\tau)<0$ indicates antibunching. Furthermore, $C_{\mu_2\mu_2^{\prime}}^{(2)}(t,t+\tau)$ has the same dimension as $G_{\mu_2\mu_2^{\prime}}^{(2)}(t,t+\tau)$, namely inverse time squared, equivalently expressed as $\rm{Hz}^2$~\cite{PhysRevLett.97.113602}.


\section{Temporal Photon Dynamics}
\label{sec:Temporal Photon Dynamics}

The scattered photons exhibit distinct dynamical behaviors under different conditions. In this section, we investigate these behaviors by analyzing the two-photon second-order correlation function Eq.~\eqref{eq:2nd correlation} together with the corresponding photon-intensity spectrum. The input pulse is chosen to be Gaussian, which is readily accessible experimentally, and the numerical parameters are chosen based on superconducting-qubit platforms. We first present and discuss the results for different ratios, and then analyze the underlying dynamical processes. 

\subsection{Dynamics under Different Lifetimes}
\label{sec:section3-1}

\begin{figure*}[t]
  \centering
  \includegraphics[width=\textwidth]{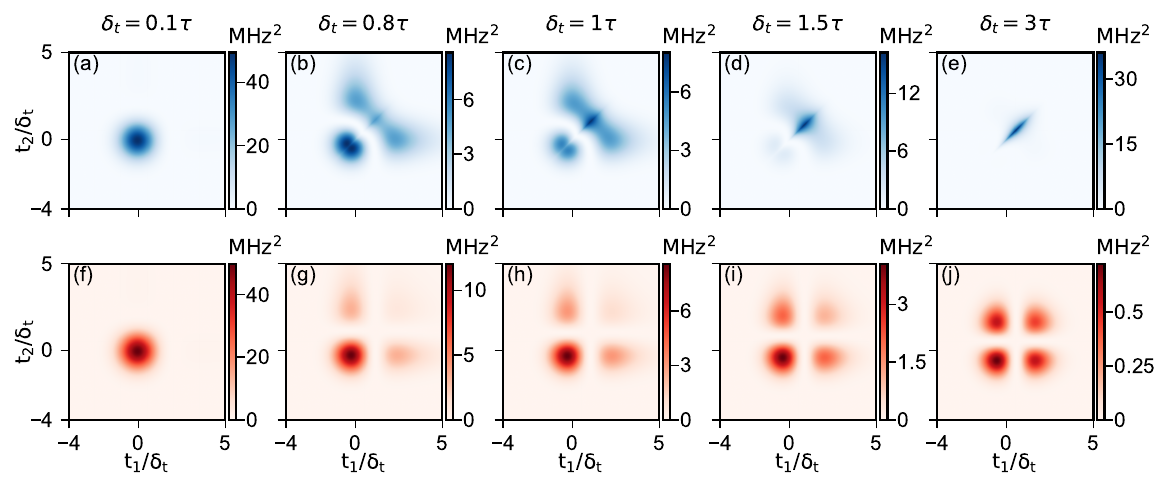}
  \caption{Second-order correlation functions of both photons are transmitted for different giant atom lifetime $\tau$ relative to the pulse width $\delta_t$. (a)-(e) Unnormalized second-order correlation function. (f)-(j) Corresponding product of the single-photon intensity spectrum.}
  \label{fig:fig_R1_g1g2_tt_resonance_coupling}
\end{figure*}

In this analysis, we vary the ratio by tuning the giant-atom lifetime while keeping the pulse width fixed, thereby allowing a direct comparison of the pulse dynamics in different regimes. Specifically, the lifetime $\tau$ is controlled through the coupling strength, while the pulse width of the Gaussian pulse is fixed at $\delta_t=100\,\rm{ns}$. The incident weak coherent pulse is centered near $5.8~\rm{GHz}$ with $\alpha=0.1$. The phase accumulated between the two coupling points is set to $\phi_0=\pi/2$, and the coupling strengths are taken to be equal and real. The detuning between the giant atom and the pulse is chosen as
\begin{equation}
\Delta = -4\pi Re(g_1g_2^*)\sin(\phi_0),
\end{equation}
so that the pulse is resonant with the giant atom. 

The results for the transmitted two-photon field are shown in Fig.~\ref{fig:fig_R1_g1g2_tt_resonance_coupling}, where the incident Gaussian is centered at $t_1=t_2=0$. In the situation illustrated in Fig.~\ref{fig:fig_R1_g1g2_tt_resonance_coupling} (a) and (f), the interaction between the giant atom and the pulse is weak. Because a Gaussian pulse in the time domain is also Gaussian in the frequency domain, its spectral width is $\delta_\omega\sim1/\delta_t$. For $\delta_t=0.1\tau$, the pulse spectrum is broad, whereas the atom interacts significantly only near resonance. As a result, the outgoing field remains nearly unchanged. 

When $\delta_t=\tau$, the interaction becomes appreciable, as shown in Figs.~\ref{fig:fig_R1_g1g2_tt_resonance_coupling} (c) and (h). The unnormalized second-order correlation function shows a bird-like pattern, while the product of the single-photon intensities evolves from a single concentrated region into four regions with different weights.

For $\delta_t=3\tau$, the interaction becomes stronger still, as shown in Fig.~\ref{fig:fig_R1_g1g2_tt_resonance_coupling} (e) and (j). In this case, the four regions in the intensity product become more uniform, while the bird-like structure in the second-order correlation function evolves into a narrow feature along $t_1=t_2$. This reflects the increasing nonlinear contribution, which is localized near $t_1=t_2$ and eventually dominates over the factorized contribution from two independent single-photon wavefunctions.

The above results are qualitatively consistent with Ref.~\cite{le2022dynamical}. Additional results for other ratios and for the three output channels are presented in Appendix~\ref{sec:Supplementary Figures of Dynamic Analysis}.


\subsection{Dynamics}
\label{sec:section3-2}

\begin{figure}
  \centering
  \includegraphics[width=1\linewidth]{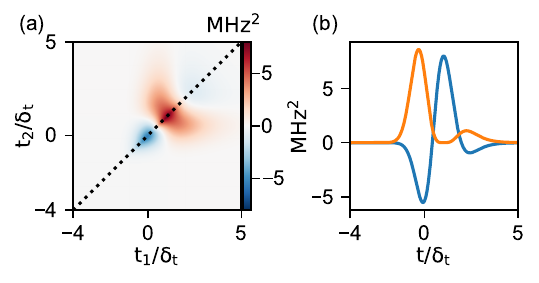}
  \caption{Transitions between bunching and antibunching in the second-order correlation function. (a) Normalized second-order correlation function. (b) Equal-time normalized second-order correlation function (blue curve, indicated by the black dashed line in panel (a)) and the product of the single-photon intensities (orange curve).}
  \label{fig:fig_R2_giant_atom_maintext1}
\end{figure}

The output spectrum discussed in the previous subsection can be understood as arising from the competition between two scattering processes, which drives transitions between bunching and antibunching in the equal-time second-order correlation function as the absolute time is varied. This behavior is shown in Fig.~\ref{fig:fig_R2_giant_atom_maintext1}, where $\phi_0=\pi/2$ and $\tau=\delta_t=100\,\rm{n s}$. All other parameters are the same as in Fig.~\ref{fig:fig_R1_g1g2_tt_resonance_coupling}. 

In Fig.~\ref{fig:fig_R2_giant_atom_maintext1}(a), the transmitted field exhibits antibunching at early times, which evolves into bunching and then crosses over to a weaker antibunching at later times. This behavior originates from two competing processes that occur when a weak coherent pulse is incident near resonance. In the first process, one photon is absorbed and re-emitted by the giant atom, while the other photon passes through without interacting. In the second process, one photon is absorbed, and the other photon forms a bound state with it, leading to their simultaneous emission.   

Because these two processes are associated with different time delays, they generate distinct temporal features, as shown in Fig.~\ref{fig:fig_R2_giant_atom_maintext1}(b). The first process corresponds to the two antibunching dips in the blue curve. The larger peak of the orange curve is associated with the photon that propagates without interacting with the atom, whereas the smaller peak corresponds to the photon that undergoes absorption and re-emission. Because these two photons occupy different temporal regions, they are unlikely to appear simultaneously, giving rise to antibunching. By contrast, the bunching peaks in the blue curve arise from the second process, in which the two photons form a bound state. This bound-state component has a larger group velocity than the single-photon component that undergoes atom-mediated scattering~\cite{tomm2023photon}, so the bunching peaks appear between the two peaks of the orange curve.

\begin{figure}
  \centering
  \includegraphics[width=1\linewidth]{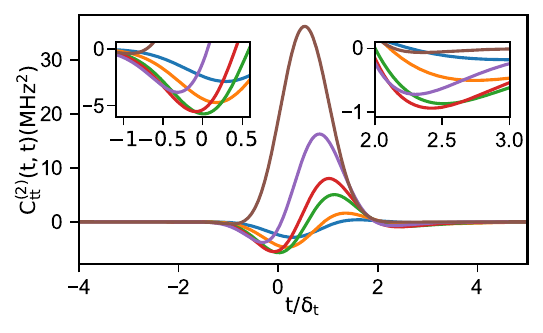}
  \caption{Normalized equal-time second-order correlation functions for different ratios of the giant atom lifetime $\tau$ to the pulse width $\delta_t$. The left and right insets zoom in on $C^{(2)}_{tt}(t,t)$ in the ranges from $-1.1$ to $0.6$ and from $2.0$ to $3.0$, respectively. The blue, orange, green, red, purple, and brown curves correspond to $\delta_t=0.3\tau,~0.5\tau,~0.8\tau,~1\tau,~1.5\tau,~3\tau$.}
  \label{fig:fig_R2_giant_atom_maintext2}
\end{figure}

Figure~\ref{fig:fig_R2_giant_atom_maintext2} shows the normalized equal-time second-order correlation function for different $\tau$ at fixed $\delta_t$, together with two insets highlighting the antibunching regions. As $\tau$ decreases, the coupling strength increases, and the bunching effect becomes progressively stronger. Stronger coupling enhances the probability of forming two-photon bound states and therefore amplifies bunching. By contrast, the antibunching first becomes stronger and then weakens. This reflects the competition between the single-photon absorption-emission process and the two-photon bound-state process: as the coupling strength increases, the former is initially enhanced, but beyond a certain threshold, the latter becomes dominant. As a result, at $\delta_t=3\tau$, the antibunching is almost completely suppressed.

\section{Phase Control}
\label{sec:Phase Control}

\begin{figure*}[t]
  \centering
  \includegraphics[width=\textwidth]{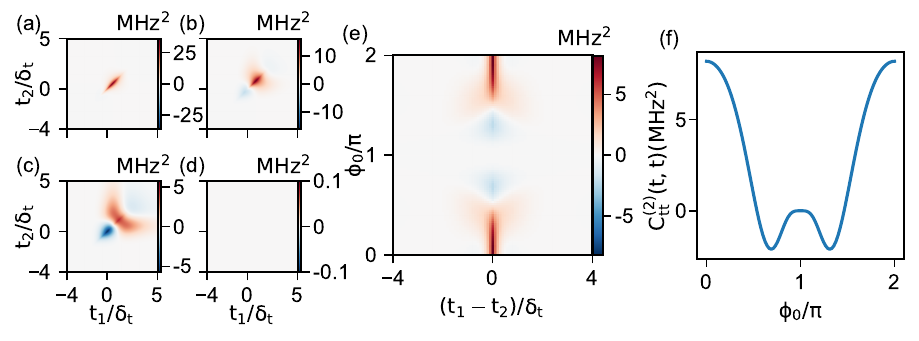}
  \caption{Normalized second-order correlation function for different phases $\phi_0$. (a) $\phi_0=0$. (b) $\phi_0=\pi/2$. (c) $\phi_0=2\pi/3$. (d) $\phi_0=\pi$. (e) Normalized second-order correlation function as a function of $\phi_0$ and $t_1-t_2$, with $t_1+t_2=-0.5\delta_t$. (f) Line cut of panel (e) at $t_1-t_2=0$, plotted as a function of $\phi_0$. }
  \label{fig:fig_R3_giant_atom_maintext}
\end{figure*}

In this section, we show that tuning the phase $\phi_0$ allows the transmitted photons to be switched among three distinct regimes—bunching, antibunching, and coherent output—each occupying a finite phase bandwidth. The results are presented in Fig.~\ref{fig:fig_R3_giant_atom_maintext}. Here $\delta_t=3\tau$ at $\phi_0=0$, and the detuning $\Delta$ is chosen to be the same as in Sec.~\ref{sec:section3-1}, so that the pulse remains resonant with the giant atom.

    Figures~\ref{fig:fig_R3_giant_atom_maintext}(a)–(d) show representative results for phases spanning half a period, from $0$ to $\pi$. At $\phi_0=0$, the condition $\delta_t=3\tau$ makes the two-photon bound-state contribution dominant, leading to behavior shown in Fig.~\ref{fig:fig_R3_giant_atom_maintext}(a), similar to that in Fig.~\ref{fig:fig_R1_g1g2_tt_resonance_coupling}(e) and (j). As $\phi_0$ increases, the effective coupling strength decreases and the giant atom lifetime becomes longer. The two scattering processes then begin to compete, and the pattern gradually evolves toward the regime corresponding to $\delta_t=\tau$, shown in Fig.~\ref{fig:fig_R3_giant_atom_maintext}(c). As $\phi_0$ approaches $\pi$, the interaction between the giant atom and the right-propagating photons gradually vanishes~\cite{PhysRevLett.120.140404}. Thus, the output field remains essentially identical to the input and retains coherent (Poissonian) statistics, as shown in Fig.~\ref{fig:fig_R3_giant_atom_maintext}(d).

Figure~\ref{fig:fig_R3_giant_atom_maintext}(e) shows a line cut along $t_1+t_2=-0.5\delta_t$, chosen so that the relevant features remain concentrated near $t_1=t_2$. As $\phi_0$ is swept over a full period $0-2\pi$, the normalized second-order correlation function evolves from bunching to antibunching, then to coherent output, back to antibunching, and finally returns to bunching. The results are symmetric about $\phi_0=\pi$. This evolution is seen more clearly in the line cut at $t_1=t_2$ shown in Fig.~\ref{fig:fig_R3_giant_atom_maintext}(f). In particular, a plateau appears near $\phi_0=\pi$, where $C_{tt}^{(2)}(t,t)=0$, indicating that the coherent regime occupies a finite bandwidth in $\phi_0$. The bunching and antibunching regimes likewise persist over finite intervals of $\phi_0$, enabling convenient phase-controlled switching among all three regimes.


\section{Discussion}
\label{sec:Discussion}
 
The model and results presented here can be implemented with superconducting circuits. In this section, we consider the feasibility by specifying realistic parameter ranges and measurement methods, and briefly discuss potential applications.  

In realistic implementations, a superconducting qubit, such as a transmon~\cite{PhysRevA.76.042319}, can work as an artificial atom~\cite{you2011atomic}. A giant atom can then be realized by coupling the qubit to a coplanar waveguide at two separated coupling points~\cite{kannan2020waveguide}. The frequency of the qubit can be tuned by varying the external flux bias and modulated in the range of  $3 -6\,\rm{GHz}$. To bring the incident field into resonance with the atom, the central frequency of the Gaussian pulse is around the artificial atom frequency, and a bias of at most $35\,\rm{MHz}$ is required. The bias is chosen according to the fixed pulse width, $\delta_t=100\rm{\mu s}$. The decay rates at the two coupling points $\Gamma_1=2\pi |g_1|^2$ and $\Gamma_2=2\pi |g_2|^2$ are varied from $0.5\,\rm{MHz}$ to $20\,\rm{MHz}$\cite{RevModPhys.93.025005}. All of these frequencies are easily achieved in superconducting platforms~\cite{RevModPhys.93.025005,krantz2019quantum}. Another important parameter is the phase accumulated between the two coupling points. The results in Section~\ref{sec:Temporal Photon Dynamics} require a fixed phase $\phi_0=\pi/2$. In Section~\ref{sec:Phase Control}, the phase is varied from $0$ to $2\pi$, which is a full period. The phase can be achieved by changing the frequency of the pulse, which is $\phi_0=k_p d, k=\omega_p/v_p$~\cite{RevModPhys.89.021001}. Another approach is to tune the complex relative phase of the coupling points $\theta_2-\theta_1$, defined as $g_1=|g_1|e^{i\theta_1},g_2=|g_2|e^{i\theta_2}$, which is a more flexible method\cite{PhysRevApplied.6.064007,roushan2017chiral,PhysRevX.13.021039,cao2024parametrically,Wang_2022}.

The second-order correlation functions can be measured using the Hanbury Brown–Twiss (HBT) method~\cite{gabelli2004hanbury}. To access the corresponding first-order correlation function, one may inject two Gaussian pulses separated by a time delay much longer than $\tau$ and then measure the second-order correlation function. In this limit, the measured signal reduces to an expression analogous to Eq.~\eqref{eq:2nd to 1st}. Related experimental protocols have been discussed in Ref.~\cite{le2022dynamical,peng2016tuneable,zhou2020tunable,hu2025demand}.

The phase accumulated between the two coupling points of a giant atom provides a tunable parameter, enabling phase-based quantum control with potential applications. As discussed in Sec.~\ref{sec:Phase Control}, in the coherent regime, especially for $\phi_0=\pi$, the output pulse remains nearly identical to the input pulse. This makes the transmitted field useful as a calibration reference for measurement and control systems~\cite {bozyigit2011antibunching,lang2013correlations,lang2014quantum}. By contrast, in the bunching and anti-bunching regimes, the outgoing pulse can be converted into a classical electrical signal by a single-photon detector. Such binary outcomes, available directly at the millikelvin stage~\cite{PhysRevX.11.011027}, may be used for low-latency control. In this way, calibration and control could be carried out using the same reference signal, thereby reducing systematic errors. 

Beyond the experimental application, several theoretical extensions are possible. For example, one may replace the Gaussian pulse with a non-Gaussian pulse or consider a sequence of Gaussian pulses with separations comparable to the atomic lifetime. In such regimes, non-Markovian effects become important~\cite{ask2022non,vodenkova2024continuous}.


\section{Conclusion}
\label{sec:Conclusion}

In conclusion, we have investigated the scattering of a weak coherent pulse incident from the left on a bidirectional waveguide coupled to a giant atom. By analyzing correlation functions, we reveal the competition between two scattering processes. One dominates at weaker coupling and leads to antibunching, while the other dominates at stronger coupling and leads to bunching. At intermediate coupling, this competition drives the equal-time second-order correlation function to switch between bunching and antibunching as the absolute time varies.

This competition is controlled by the relevant ratio and can be further tuned through the phase accumulated between two coupling points of the giant atom. Consequently, over a full phase period, transmitted photons can be switched among three different regimes--bunching, antibunching, and coherent output--each extending over a finite phase interval. This provides a convenient mechanism for phase-controlled switching and suggests potential applications in quantum control.

We have further discussed the feasibility of observing these effects in the superconducting platform, including realistic parameter ranges and measurement protocols. More broadly, our work provides a route to controlling photon pulses with giant atoms and understanding the resulting photon-correlation dynamics. It also opens the door to future studies of multi-giant-atom systems, as well as non-Markovian effects arising from multiple incident pulses.


\section{Acknowledgments}

This work was supported by the National Natural Science Foundation of China (Grant No. 12375025).

\bibliography{references_with_urls}

@phdthesis{lang2014quantum,
  title={Quantum microwave radiation and its interference characterized by correlation function measurements in circuit quantum electrodynamics},
  author={Lang, Christian},
  year={2014},
  school={ETH Zurich}
}

@article{lang2013correlations,
  title={Correlations, indistinguishability and entanglement in Hong--Ou--Mandel experiments at microwave frequencies},
  author={Lang, C and Eichler, Christopher and Steffen, L and Fink, JM and Woolley, Matthew J and Blais, Alexandre and Wallraff, Andreas},
  journal={Nature Physics},
  volume={9},
  number={6},
  pages={345--348},
  year={2013},
  publisher={Nature Publishing Group UK London},
  url={https://www.nature.com/articles/nphys2612}
}

@article{bozyigit2011antibunching,
  title={Antibunching of microwave-frequency photons observed in correlation measurements using linear detectors},
  author={Bozyigit, Deniz and Lang, C and Steffen, L and Fink, JM and Eichler, Christopher and Baur, M and Bianchetti, R and Leek, Peter J and Filipp, Stefan and Da Silva, Marcus P and others},
  journal={Nature Physics},
  volume={7},
  number={2},
  pages={154--158},
  year={2011},
  publisher={Nature Publishing Group UK London},
  url={https://www.nature.com/articles/nphys1845}
}

@article{vodenkova2024continuous,
  title={Continuous coherent quantum feedback with time delays: Tensor network solution},
  author={Vodenkova, Kseniia and Pichler, Hannes},
  journal={Physical Review X},
  volume={14},
  number={3},
  pages={031043},
  year={2024},
  publisher={APS},
  url={https://link.aps.org/doi/10.1103/PhysRevX.14.031043}
}

@article{ask2022non,
  title={Non-Markovian steady states of a driven two-level system},
  author={Ask, Andreas and Johansson, G{\"o}ran},
  journal={Physical Review Letters},
  volume={128},
  number={8},
  pages={083603},
  year={2022},
  publisher={APS},
  url={https://link.aps.org/doi/10.1103/PhysRevLett.128.083603}
}

@article{hu2025demand,
  title={On-Demand Microwave Single-Photon Source Based on Tantalum Thin Film},
  author={Hu, Ying and Li, Sheng-Yong and Chen, En-Qi and Zhang, Jing and Liu, Yu-xi and Feng, Jia-Gui and Peng, Zhihui},
  journal={arXiv preprint arXiv:2512.07589},
  year={2025},
  url={https://arxiv.org/abs/2512.07589}
}

@article{zhou2020tunable,
  title={Tunable microwave single-photon source based on transmon qubit with high efficiency},
  author={Zhou, Yu and Peng, Zhihui and Horiuchi, Yuta and Astafiev, OV and Tsai, JS},
  journal={Physical Review Applied},
  volume={13},
  number={3},
  pages={034007},
  year={2020},
  publisher={APS},
  url={https://link.aps.org/doi/10.1103/PhysRevApplied.13.034007}
}

@article{gabelli2004hanbury,
  title={Hanbury Brown--Twiss correlations to probe the population statistics of GHz photons emitted by conductors},
  author={Gabelli, Julien and Reydellet, L-H and Feve, Gwendal and Berroir, J-M and Placais, Bernard and Roche, Patrice and Glattli, D Christian},
  journal={Physical review letters},
  volume={93},
  number={5},
  pages={056801},
  year={2004},
  publisher={APS},
  url={https://link.aps.org/doi/10.1103/PhysRevLett.93.056801}
}

@article{peng2016tuneable,
  title={Tuneable on-demand single-photon source in the microwave range},
  author={Peng, ZH and De Graaf, SE and Tsai, JS and Astafiev, OV},
  journal={Nature communications},
  volume={7},
  number={1},
  pages={12588},
  year={2016},
  publisher={Nature Publishing Group UK London},
  url={https://www.nature.com/articles/ncomms12588}
}

@article{roushan2017chiral,
  title={Chiral ground-state currents of interacting photons in a synthetic magnetic field},
  author={Roushan, Pedram and Neill, Charles and Megrant, Anthony and Chen, Yu and Babbush, Ryan and Barends, Rami and Campbell, Brooks and Chen, Zijun and Chiaro, Ben and Dunsworth, Andrew and others},
  journal={Nature Physics},
  volume={13},
  number={2},
  pages={146--151},
  year={2017},
  publisher={Nature Publishing Group UK London},
  url={https://www.nature.com/articles/nphys3930}
}

@article{cao2024parametrically,
  title={Parametrically controlled chiral interface for superconducting quantum devices},
  author={Cao, Xi and Irfan, Abdullah and Mollenhauer, Michael and Singirikonda, Kaushik and Pfaff, Wolfgang},
  journal={Physical Review Applied},
  volume={22},
  number={6},
  pages={064023},
  year={2024},
  publisher={APS},
  url={https://link.aps.org/doi/10.1103/PhysRevApplied.22.064023}
}

@article{Wang_2022,
doi = {10.1088/2058-9565/ac6a04},
url = {https://dx.doi.org/10.1088/2058-9565/ac6a04},
year = {2022},
month = {may},
publisher = {IOP Publishing},
volume = {7},
number = {3},
pages = {035007},
author = {Xin Wang and Hong-Rong Li},
title = {Chiral quantum network with giant atoms},
journal = {Quantum Science and Technology}
}

@article{PhysRevApplied.6.064007,
  title = {Universal Gate for Fixed-Frequency Qubits via a Tunable Bus},
  author = {McKay, David C. and Filipp, Stefan and Mezzacapo, Antonio and Magesan, Easwar and Chow, Jerry M. and Gambetta, Jay M.},
  journal = {Phys. Rev. Appl.},
  volume = {6},
  issue = {6},
  pages = {064007},
  numpages = {10},
  year = {2016},
  month = {Dec},
  publisher = {American Physical Society},
  doi = {10.1103/PhysRevApplied.6.064007},
  url = {https://link.aps.org/doi/10.1103/PhysRevApplied.6.064007}
}

@article{RevModPhys.89.021001,
  title = {Colloquium: Strongly interacting photons in one-dimensional continuum},
  author = {Roy, Dibyendu and Wilson, C. M. and Firstenberg, Ofer},
  journal = {Rev. Mod. Phys.},
  volume = {89},
  issue = {2},
  pages = {021001},
  numpages = {23},
  year = {2017},
  month = {May},
  publisher = {American Physical Society},
  doi = {10.1103/RevModPhys.89.021001},
  url = {https://link.aps.org/doi/10.1103/RevModPhys.89.021001}
}

@article{krantz2019quantum,
  title={A quantum engineer's guide to superconducting qubits},
  author={Krantz, Philip and Kjaergaard, Morten and Yan, Fei and Orlando, Terry P and Gustavsson, Simon and Oliver, William D},
  journal={Applied physics reviews},
  volume={6},
  number={2},
  year={2019},
  publisher={AIP Publishing},
  url={https://pubs.aip.org/aip/apr/article/6/2/021318/570326/A-quantum-engineer-s-guide-to-superconducting}
}

@article{RevModPhys.95.015002,
  title = {Waveguide quantum electrodynamics: Collective radiance and photon-photon correlations},
  author = {Sheremet, Alexandra S. and Petrov, Mihail I. and Iorsh, Ivan V. and Poshakinskiy, Alexander V. and Poddubny, Alexander N.},
  journal = {Rev. Mod. Phys.},
  volume = {95},
  issue = {1},
  pages = {015002},
  numpages = {59},
  year = {2023},
  month = {Mar},
  publisher = {American Physical Society},
  doi = {10.1103/RevModPhys.95.015002},
  url = {https://link.aps.org/doi/10.1103/RevModPhys.95.015002}
}

@article{PhysRevLett.98.153003,
  title = {Strongly Correlated Two-Photon Transport in a One-Dimensional Waveguide Coupled to a Two-Level System},
  author = {Shen, Jung-Tsung and Fan, Shanhui},
  journal = {Phys. Rev. Lett.},
  volume = {98},
  issue = {15},
  pages = {153003},
  numpages = {4},
  year = {2007},
  month = {Apr},
  publisher = {American Physical Society},
  doi = {10.1103/PhysRevLett.98.153003},
  url = {https://link.aps.org/doi/10.1103/PhysRevLett.98.153003}
}

@article{lund2024subtraction,
  title={Subtraction and addition of propagating photons by two-level emitters},
  author={Lund, Mads M and Yang, Fan and Christiansen, Victor Rueskov and Kornovan, Danil and M{\o}lmer, Klaus},
  journal={Physical Review Letters},
  volume={133},
  number={10},
  pages={103601},
  year={2024},
  publisher={APS},
  url={https://link.aps.org/doi/10.1103/PhysRevLett.133.103601}
}

@article{tian2025disorder,
  title={Disorder-Induced Strongly Correlated Photons in Waveguide QED},
  author={Tian, Guoqing and Zheng, Li-Li and Zhan, Zhi-Ming and Nori, Franco and L{\"u}, Xin-You},
  journal={Physical Review Letters},
  volume={135},
  number={15},
  pages={153604},
  year={2025},
  publisher={APS},
  url = {https://link.aps.org/doi/10.1103/mldt-d59t}
}

@article{le2021experimental,
  title={Experimental reconstruction of the few-photon nonlinear scattering matrix from a single quantum dot in a nanophotonic waveguide},
  author={Le Jeannic, Hanna and Ramos, Tom{\'a}s and Simonsen, Signe F and Pregnolato, Tommaso and Liu, Zhe and Schott, R{\"u}diger and Wieck, Andreas D and Ludwig, Arne and Rotenberg, Nir and Garc{\'\i}a-Ripoll, Juan Jos{\'e} and others},
  journal={Physical Review Letters},
  volume={126},
  number={2},
  pages={023603},
  year={2021},
  publisher={APS},
  url={https://link.aps.org/doi/10.1103/PhysRevLett.126.023603}
}

@article{le2022dynamical,
  title={Dynamical photon--photon interaction mediated by a quantum emitter},
  author={Le Jeannic, Hanna and Tiranov, Alexey and Carolan, Jacques and Ramos, Tom{\'a}s and Wang, Ying and Appel, Martin Hayhurst and Scholz, Sven and Wieck, Andreas D and Ludwig, Arne and Rotenberg, Nir and others},
  journal={Nature Physics},
  volume={18},
  number={10},
  pages={1191--1195},
  year={2022},
  publisher={Nature Publishing Group UK London},
  url={https://www.nature.com/articles/s41567-022-01720-x}
}

@article{kannan2020waveguide,
  title={Waveguide quantum electrodynamics with superconducting artificial giant atoms},
  author={Kannan, Bharath and Ruckriegel, Max J and Campbell, Daniel L and Frisk Kockum, Anton and Braum{\"u}ller, Jochen and Kim, David K and Kjaergaard, Morten and Krantz, Philip and Melville, Alexander and Niedzielski, Bethany M and others},
  journal={Nature},
  volume={583},
  number={7818},
  pages={775--779},
  year={2020},
  publisher={Nature Publishing Group UK London},
  url={https://www.nature.com/articles/s41586-020-2529-9}
}

@article{wang2022giant,
  title={Giant spin ensembles in waveguide magnonics},
  author={Wang, Zi-Qi and Wang, Yi-Pu and Yao, Jiguang and Shen, Rui-Chang and Wu, Wei-Jiang and Qian, Jie and Li, Jie and Zhu, Shi-Yao and You, JQ},
  journal={Nature communications},
  volume={13},
  number={1},
  pages={7580},
  year={2022},
  publisher={Nature Publishing Group UK London},
  url={https://www.nature.com/articles/s41467-022-35174-9}
}

@article{chen2022nonreciprocal,
  title={Nonreciprocal and chiral single-photon scattering for giant atoms},
  author={Chen, Yao-Tong and Du, Lei and Guo, Lingzhen and Wang, Zhihai and Zhang, Yan and Li, Yong and Wu, Jin-Hui},
  journal={Communications Physics},
  volume={5},
  number={1},
  pages={215},
  year={2022},
  publisher={Nature Publishing Group UK London},
  url={https://www.nature.com/articles/s42005-022-00991-3}
}

@article{PhysRevA.105.023712,
  title = {Chiral quantum optics with giant atoms},
  author = {Soro, Ariadna and Kockum, Anton Frisk},
  journal = {Phys. Rev. A},
  volume = {105},
  issue = {2},
  pages = {023712},
  numpages = {16},
  year = {2022},
  month = {Feb},
  publisher = {American Physical Society},
  doi = {10.1103/PhysRevA.105.023712},
  url = {https://link.aps.org/doi/10.1103/PhysRevA.105.023712}
}

@article{PhysRevLett.133.063603,
  title = {Controlling Markovianity with Chiral Giant Atoms},
  author = {Roccati, Federico and Cilluffo, Dario},
  journal = {Phys. Rev. Lett.},
  volume = {133},
  issue = {6},
  pages = {063603},
  numpages = {6},
  year = {2024},
  month = {Aug},
  publisher = {American Physical Society},
  doi = {10.1103/PhysRevLett.133.063603},
  url = {https://link.aps.org/doi/10.1103/PhysRevLett.133.063603}
}

@article{PhysRevLett.120.140404,
  title = {Decoherence-Free Interaction between Giant Atoms in Waveguide Quantum Electrodynamics},
  author = {Kockum, Anton Frisk and Johansson, G\"oran and Nori, Franco},
  journal = {Phys. Rev. Lett.},
  volume = {120},
  issue = {14},
  pages = {140404},
  numpages = {8},
  year = {2018},
  month = {Apr},
  publisher = {American Physical Society},
  doi = {10.1103/PhysRevLett.120.140404},
  url = {https://link.aps.org/doi/10.1103/PhysRevLett.120.140404}
}

@article{PhysRevA.107.023705,
  title = {Complex decoherence-free interactions between giant atoms},
  author = {Du, Lei and Guo, Lingzhen and Li, Yong},
  journal = {Phys. Rev. A},
  volume = {107},
  issue = {2},
  pages = {023705},
  numpages = {11},
  year = {2023},
  month = {Feb},
  publisher = {American Physical Society},
  doi = {10.1103/PhysRevA.107.023705},
  url = {https://link.aps.org/doi/10.1103/PhysRevA.107.023705}
}

@article{PhysRevResearch.2.043184,
  title = {Mechanism of decoherence-free coupling between giant atoms},
  author = {Carollo, Angelo and Cilluffo, Dario and Ciccarello, Francesco},
  journal = {Phys. Rev. Res.},
  volume = {2},
  issue = {4},
  pages = {043184},
  numpages = {13},
  year = {2020},
  month = {Nov},
  publisher = {American Physical Society},
  doi = {10.1103/PhysRevResearch.2.043184},
  url = {https://link.aps.org/doi/10.1103/PhysRevResearch.2.043184}
}

@book{Scully_Zubairy_1997, 
    place={Cambridge}, 
    title={Quantum Optics}, 
    publisher={Cambridge University Press}, 
    author={Scully, Marlan O. and Zubairy, M. Suhail}, 
    year={1997}
}

@book{agarwal2012quantum,
  title={Quantum optics},
  author={Agarwal, Girish S},
  year={2012},
  publisher={Cambridge University Press}
}

@article{RevModPhys.73.565,
  title = {Manipulating quantum entanglement with atoms and photons in a cavity},
  author = {Raimond, J. M. and Brune, M. and Haroche, S.},
  journal = {Rev. Mod. Phys.},
  volume = {73},
  issue = {3},
  pages = {565--582},
  numpages = {0},
  year = {2001},
  month = {Aug},
  publisher = {American Physical Society},
  doi = {10.1103/RevModPhys.73.565},
  url = {https://link.aps.org/doi/10.1103/RevModPhys.73.565}
}

@article{bin2018two,
  title={Two-photon blockade in a cascaded cavity-quantum-electrodynamics system},
  author={Bin, Qian and L{\"u}, Xin-You and Bin, Shang-Wu and Wu, Ying},
  journal={Physical Review A},
  volume={98},
  number={4},
  pages={043858},
  year={2018},
  publisher={APS},
  url={https://link.aps.org/doi/10.1103/PhysRevA.98.043858}
}

@article{lu2025chiral,
  title={Chiral interaction induced near-perfect photon blockade},
  author={Lu, Zhi-Guang and Wu, Ying and L{\"u}, Xin-You},
  journal={Physical Review Letters},
  volume={134},
  number={1},
  pages={013602},
  year={2025},
  publisher={APS},
  url={https://link.aps.org/doi/10.1103/PhysRevLett.134.013602}
}

@article{mahmoodian2020dynamics,
  title={Dynamics of many-body photon bound states in chiral waveguide QED},
  author={Mahmoodian, Sahand and Calaj{\'o}, Giuseppe and Chang, Darrick E and Hammerer, Klemens and S{\o}rensen, Anders S},
  journal={Physical Review X},
  volume={10},
  number={3},
  pages={031011},
  year={2020},
  publisher={APS},
  url={https://link.aps.org/doi/10.1103/PhysRevX.10.031011}
}

@article{gu2023correlated,
  title={Correlated two-photon scattering in a one-dimensional waveguide coupled to two-or three-level giant atoms},
  author={Gu, Wenju and Huang, He and Yi, Zhen and Chen, Lei and Sun, Lihui and Tan, Huatang},
  journal={Physical Review A},
  volume={108},
  number={5},
  pages={053718},
  year={2023},
  publisher={APS},
  url={https://link.aps.org/doi/10.1103/PhysRevA.108.053718}
}

@article{PhysRevA.88.043806,
  title = {Input-output theory for waveguide QED with an ensemble of inhomogeneous atoms},
  author = {Lalumi\`ere, Kevin and Sanders, Barry C. and van Loo, A. F. and Fedorov, A. and Wallraff, A. and Blais, A.},
  journal = {Phys. Rev. A},
  volume = {88},
  issue = {4},
  pages = {043806},
  numpages = {15},
  year = {2013},
  month = {Oct},
  publisher = {American Physical Society},
  doi = {10.1103/PhysRevA.88.043806},
  url = {https://link.aps.org/doi/10.1103/PhysRevA.88.043806}
}

@article{PhysRevA.93.013828,
  title = {Green's-function formalism for waveguide QED applications},
  author = {Schneider, Michael P. and Sproll, Tobias and Stawiarski, Christina and Schmitteckert, Peter and Busch, Kurt},
  journal = {Phys. Rev. A},
  volume = {93},
  issue = {1},
  pages = {013828},
  numpages = {17},
  year = {2016},
  month = {Jan},
  publisher = {American Physical Society},
  doi = {10.1103/PhysRevA.93.013828},
  url = {https://link.aps.org/doi/10.1103/PhysRevA.93.013828}
}

@article{PhysRevA.108.053703,
  title = {Analytical approach to higher-order correlation functions in U(1) symmetric systems},
  author = {Lu, Zhi-Guang and Shang, Cheng and Wu, Ying and L\"u, Xin-You},
  journal = {Phys. Rev. A},
  volume = {108},
  issue = {5},
  pages = {053703},
  numpages = {20},
  year = {2023},
  month = {Nov},
  publisher = {American Physical Society},
  doi = {10.1103/PhysRevA.108.053703},
  url = {https://link.aps.org/doi/10.1103/PhysRevA.108.053703}
}

@article{ke2019inelastic,
  title={Inelastic scattering of photon pairs in qubit arrays with subradiant states},
  author={Ke, Yongguan and Poshakinskiy, Alexander V and Lee, Chaohong and Kivshar, Yuri S and Poddubny, Alexander N},
  journal={Physical review letters},
  volume={123},
  number={25},
  pages={253601},
  year={2019},
  publisher={APS},
  url={https://link.aps.org/doi/10.1103/PhysRevLett.123.253601}
}

@article{PhysRevA.84.063803,
  title = {Two-photon transport in a waveguide coupled to a cavity in a two-level system},
  author = {Shi, T. and Fan, Shanhui and Sun, C. P.},
  journal = {Phys. Rev. A},
  volume = {84},
  issue = {6},
  pages = {063803},
  numpages = {8},
  year = {2011},
  month = {Dec},
  publisher = {American Physical Society},
  doi = {10.1103/PhysRevA.84.063803},
  url = {https://link.aps.org/doi/10.1103/PhysRevA.84.063803}
}

@article{PhysRevA.85.015803,
  title = {Control of correlated two-photon transport in a one-dimensional waveguide},
  author = {Yan, Wei-Bin and Fan, Qiu-Bo and Zhou, Ling},
  journal = {Phys. Rev. A},
  volume = {85},
  issue = {1},
  pages = {015803},
  numpages = {4},
  year = {2012},
  month = {Jan},
  publisher = {American Physical Society},
  doi = {10.1103/PhysRevA.85.015803},
  url = {https://link.aps.org/doi/10.1103/PhysRevA.85.015803}
}

@article{nysteen2015scattering,
  title={Scattering of two photons on a quantum emitter in a one-dimensional waveguide: exact dynamics and induced correlations},
  author={Nysteen, Anders and Kristensen, Philip Tr{\o}st and McCutcheon, Dara PS and Kaer, Per and M{\o}rk, Jesper},
  journal={New Journal of Physics},
  volume={17},
  number={2},
  pages={023030},
  year={2015},
  publisher={IOP Publishing},
  url={https://doi.org/10.1088/1367-2630/17/2/023030}
}

@article{fan2010input,
  title={Input-output formalism for few-photon transport in one-dimensional nanophotonic waveguides coupled to a qubit},
  author={Fan, Shanhui and Kocaba{\c{s}}, {\c{S}}{\"u}kr{\"u} Ekin and Shen, Jung-Tsung},
  journal={Physical Review A—Atomic, Molecular, and Optical Physics},
  volume={82},
  number={6},
  pages={063821},
  year={2010},
  publisher={APS},
  url={https://link.aps.org/doi/10.1103/PhysRevA.82.063821}
}

@article{kubanek2008two,
  title={Two-photon gateway in one-atom cavity quantum electrodynamics},
  author={Kubanek, Alexander and Ourjoumtsev, Alexei and Schuster, Ingrid and Koch, Markus and Pinkse, Pepijn WH and Murr, Karim and Rempe, Gerhard},
  journal={Physical Review Letters},
  volume={101},
  number={20},
  pages={203602},
  year={2008},
  publisher={APS},
  url={https://link.aps.org/doi/10.1103/PhysRevLett.101.203602}
}

@article{gulfam2018highly,
  title={Highly directional photon superbunching from a few-atom chain of emitters},
  author={Gulfam, Qurrat-ul-Ain and Ficek, Zbigniew},
  journal={Physical Review A},
  volume={98},
  number={6},
  pages={063824},
  year={2018},
  publisher={APS},
  url={https://link.aps.org/doi/10.1103/PhysRevA.98.063824}
}

@article{tomm2023photon,
  title={Photon bound state dynamics from a single artificial atom},
  author={Tomm, Natasha and Mahmoodian, Sahand and Antoniadis, Nadia O and Schott, R{\"u}diger and Valentin, Sascha R and Wieck, Andreas D and Ludwig, Arne and Javadi, Alisa and Warburton, Richard J},
  journal={Nature Physics},
  volume={19},
  number={6},
  pages={857--862},
  year={2023},
  publisher={Nature Publishing Group UK London},
  url={https://www.nature.com/articles/s41567-023-01997-6}
}

@article{gong2024tunable,
  title={Tunable quantum router with giant atoms, implementing quantum gates, teleportation, non-reciprocity, and circulators},
  author={Gong, Rui-Yang and He, Zi-Yu and Yu, Cheng-He and Zhang, Ge-Fei and Nori, Franco and Xiang, Ze-Liang},
  journal={arXiv preprint arXiv:2411.19307},
  year={2024},
  url={https://arxiv.org/abs/2411.19307}
}

@article{PhysRevA.111.033716,
  title = {Dark-state-mediated topological response to coherence effects from two independent phases in single-photon transport},
  author = {Zhou, Haohang and Wang, Luojia and Gong, Rui-Yang and Xiang, Ze-Liang and Chen, Xianfeng and Yuan, Luqi},
  journal = {Phys. Rev. A},
  volume = {111},
  issue = {3},
  pages = {033716},
  numpages = {7},
  year = {2025},
  month = {Mar},
  publisher = {American Physical Society},
  doi = {10.1103/PhysRevA.111.033716},
  url = {https://link.aps.org/doi/10.1103/PhysRevA.111.033716}
}

@article{ylrq-98wl,
  title = {Controllable operations of edge states in cross-one-dimensional topological chains},
  author = {Lu, Xian-Liang and Xiang, Ze-Liang},
  journal = {Phys. Rev. Res.},
  volume = {7},
  issue = {4},
  pages = {L042050},
  numpages = {7},
  year = {2025},
  month = {Dec},
  publisher = {American Physical Society},
  doi = {10.1103/ylrq-98wl},
  url = {https://link.aps.org/doi/10.1103/ylrq-98wl}
}

@article{PhysRevX.11.011027,
  title = {High-Fidelity Measurement of a Superconducting Qubit Using an On-Chip Microwave Photon Counter},
  author = {Opremcak, A. and Liu, C. H. and Wilen, C. and Okubo, K. and Christensen, B. G. and Sank, D. and White, T. C. and Vainsencher, A. and Giustina, M. and Megrant, A. and Burkett, B. and Plourde, B. L. T. and McDermott, R.},
  journal = {Phys. Rev. X},
  volume = {11},
  issue = {1},
  pages = {011027},
  numpages = {15},
  year = {2021},
  month = {Feb},
  publisher = {American Physical Society},
  doi = {10.1103/PhysRevX.11.011027},
  url = {https://link.aps.org/doi/10.1103/PhysRevX.11.011027}
}

@article{rosenblum2016extraction,
  title={Extraction of a single photon from an optical pulse},
  author={Rosenblum, Serge and Bechler, Orel and Shomroni, Itay and Lovsky, Yulia and Guendelman, Gabriel and Dayan, Barak},
  journal={Nature Photonics},
  volume={10},
  number={1},
  pages={19--22},
  year={2016},
  publisher={Nature Publishing Group UK London},
  url={https://www.nature.com/articles/nphoton.2015.227}
}

@article{PhysRevLett.121.143601,
  title = {Strongly Correlated Photon Transport in Waveguide Quantum Electrodynamics with Weakly Coupled Emitters},
  author = {Mahmoodian, Sahand and \ifmmode \check{C}\else \v{C}\fi{}epulkovskis, Mantas and Das, Sumanta and Lodahl, Peter and Hammerer, Klemens and S\o{}rensen, Anders S.},
  journal = {Phys. Rev. Lett.},
  volume = {121},
  issue = {14},
  pages = {143601},
  numpages = {6},
  year = {2018},
  month = {Oct},
  publisher = {American Physical Society},
  doi = {10.1103/PhysRevLett.121.143601},
  url = {https://link.aps.org/doi/10.1103/PhysRevLett.121.143601}
}

@phdthesis{vcepulkovskis2017nonlinear,
  title={Nonlinear photon interactions in waveguides},
  author={{\v{C}}epulkovskis, Mantas},
  year={2017}
}

@article{poshakinskiy2021quantum,
  title={Quantum Hall phases emerging from atom--photon interactions},
  author={Poshakinskiy, Alexander V and Zhong, Janet and Ke, Yongguan and Olekhno, Nikita A and Lee, Chaohong and Kivshar, Yuri S and Poddubny, Alexander N},
  journal={npj Quantum Information},
  volume={7},
  number={1},
  pages={34},
  year={2021},
  publisher={Nature Publishing Group UK London},
  url={https://www.nature.com/articles/s41534-021-00372-8}
}

@article{prasad2020correlating,
  title={Correlating photons using the collective nonlinear response of atoms weakly coupled to an optical mode},
  author={Prasad, Adarsh S and Hinney, Jakob and Mahmoodian, Sahand and Hammerer, Klemens and Rind, Samuel and Schneeweiss, Philipp and S{\o}rensen, Anders S and Volz, J{\"u}rgen and Rauschenbeutel, Arno},
  journal={Nature Photonics},
  volume={14},
  number={12},
  pages={719--722},
  year={2020},
  publisher={Nature Publishing Group UK London},
  url={https://www.nature.com/articles/s41566-020-0692-z}
}

@inproceedings{kockum2021quantum,
  title={Quantum optics with giant atoms—the first five years},
  author={Kockum, A Frisk},
  booktitle={International Symposium on Mathematics, Quantum Theory, and Cryptography},
  volume={33},
  pages={125--146},
  year={2021},
  organization={Springer Singapore}
}

@article{andersson2019non,
  title={Non-exponential decay of a giant artificial atom},
  author={Andersson, Gustav and Suri, Baladitya and Guo, Lingzhen and Aref, Thomas and Delsing, Per},
  journal={Nature Physics},
  volume={15},
  number={11},
  pages={1123--1127},
  year={2019},
  publisher={Nature Publishing Group UK London},
  url={https://www.nature.com/articles/s41567-019-0605-6}
}

@article{
doi:10.1126/science.1257219,
author = {Martin V. Gustafsson  and Thomas Aref  and Anton Frisk Kockum  and Maria K. Ekström  and Göran Johansson  and Per Delsing },
title = {Propagating phonons coupled to an artificial atom},
journal = {Science},
volume = {346},
number = {6206},
pages = {207-211},
year = {2014},
doi = {10.1126/science.1257219},
URL = {https://www.science.org/doi/abs/10.1126/science.1257219}}

@article{qiu2023collective,
  title={Collective radiance of giant atoms in non-Markovian regime},
  author={Qiu, Qing-Yang and Wu, Ying and L{\"u}, Xin-You},
  journal={Science China Physics, Mechanics \& Astronomy},
  volume={66},
  number={2},
  pages={224212},
  year={2023},
  publisher={Springer},
  url={https://link.springer.com/article/10.1007/s11433-022-1990-x}
}

@article{PhysRevLett.122.243602,
  title = {Photon Blockade in Weakly Driven Cavity Quantum Electrodynamics Systems with Many Emitters},
  author = {Trivedi, Rahul and Radulaski, Marina and Fischer, Kevin A. and Fan, Shanhui and Vu\ifmmode \check{c}\else \v{c}\fi{}kovi\ifmmode \acute{c}\else \'{c}\fi{}, Jelena},
  journal = {Phys. Rev. Lett.},
  volume = {122},
  issue = {24},
  pages = {243602},
  numpages = {6},
  year = {2019},
  month = {Jun},
  publisher = {American Physical Society},
  doi = {10.1103/PhysRevLett.122.243602},
  url = {https://link.aps.org/doi/10.1103/PhysRevLett.122.243602}
}

@article{liang2018observation,
  title={Observation of three-photon bound states in a quantum nonlinear medium},
  author={Liang, Qi-Yu and Venkatramani, Aditya V and Cantu, Sergio H and Nicholson, Travis L and Gullans, Michael J and Gorshkov, Alexey V and Thompson, Jeff D and Chin, Cheng and Lukin, Mikhail D and Vuleti{\'c}, Vladan},
  journal={Science},
  volume={359},
  number={6377},
  pages={783--786},
  year={2018},
  publisher={American Association for the Advancement of Science},
  url={https://www.science.org/doi/10.1126/science.aao7293}
}

@article{walther2006cavity,
  title={Cavity quantum electrodynamics},
  author={Walther, Herbert and Varcoe, Benjamin TH and Englert, Berthold-Georg and Becker, Thomas},
  journal={Reports on Progress in Physics},
  volume={69},
  number={5},
  pages={1325--1382},
  year={2006},
  url={https://doi.org/10.1088/0034-4885/69/5/R02}
}

@article{PhysRevA.93.062104,
  title = {Non-Markovian dynamics in chiral quantum networks with spins and photons},
  author = {Ramos, Tom\'as and Vermersch, Beno\^{\i}t and Hauke, Philipp and Pichler, Hannes and Zoller, Peter},
  journal = {Phys. Rev. A},
  volume = {93},
  issue = {6},
  pages = {062104},
  numpages = {23},
  year = {2016},
  month = {Jun},
  publisher = {American Physical Society},
  doi = {10.1103/PhysRevA.93.062104},
  url = {https://link.aps.org/doi/10.1103/PhysRevA.93.062104}
}

@article{daley2014quantum,
  title={Quantum trajectories and open many-body quantum systems},
  author={Daley, Andrew J},
  journal={Advances in Physics},
  volume={63},
  number={2},
  pages={77--149},
  year={2014},
  publisher={Taylor \& Francis},
  url={https://www.tandfonline.com/doi/abs/10.1080/00018732.2014.933502}
}

@article{PhysRevX.5.041017,
  title = {Photon Temporal Modes: A Complete Framework for Quantum Information Science},
  author = {Brecht, B. and Reddy, Dileep V. and Silberhorn, C. and Raymer, M. G.},
  journal = {Phys. Rev. X},
  volume = {5},
  issue = {4},
  pages = {041017},
  numpages = {17},
  year = {2015},
  month = {Oct},
  publisher = {American Physical Society},
  doi = {10.1103/PhysRevX.5.041017},
  url = {https://link.aps.org/doi/10.1103/PhysRevX.5.041017}
}

@article{PhysRevA.90.013837,
  title = {Designing frequency-dependent relaxation rates and Lamb shifts for a giant artificial atom},
  author = {Frisk Kockum, Anton and Delsing, Per and Johansson, G\"oran},
  journal = {Phys. Rev. A},
  volume = {90},
  issue = {1},
  pages = {013837},
  numpages = {13},
  year = {2014},
  month = {Jul},
  publisher = {American Physical Society},
  doi = {10.1103/PhysRevA.90.013837},
  url = {https://link.aps.org/doi/10.1103/PhysRevA.90.013837}
}

@article{huang2022exceptional,
  title={Exceptional Photon Blockade: Engineering Photon Blockade with Chiral Exceptional Points (Laser Photonics Rev. 16 (7)/2022)},
  author={Huang, Ran and {\"O}zdemir, {\c{S}}K and Liao, Jie-Qiao and Minganti, Fabrizio and Kuang, Le-Man and Nori, Franco and Jing, Hui},
  journal={Laser \& Photonics Reviews},
  volume={16},
  number={7},
  pages={2270034},
  year={2022},
  publisher={Wiley Online Library},
  url={https://onlinelibrary.wiley.com/doi/abs/10.1002/lpor.202270034}
}

@inproceedings{frisk2017quantum,
  title={Quantum optics with giant artificial atoms in a 1D waveguide},
  author={Frisk Kockum, Anton and Johansson, G{\"o}ran and Nori, Franco},
  booktitle={APS March Meeting Abstracts},
  volume={2017},
  pages={R46--010},
  year={2017}
}

@article{PhysRevLett.126.043602,
  title = {Tunable Chiral Bound States with Giant Atoms},
  author = {Wang, Xin and Liu, Tao and Kockum, Anton Frisk and Li, Hong-Rong and Nori, Franco},
  journal = {Phys. Rev. Lett.},
  volume = {126},
  issue = {4},
  pages = {043602},
  numpages = {8},
  year = {2021},
  month = {Jan},
  publisher = {American Physical Society},
  doi = {10.1103/PhysRevLett.126.043602},
  url = {https://link.aps.org/doi/10.1103/PhysRevLett.126.043602}
}

@article{yan2025giant,
  title={Giant-Atom Quantum Batteries},
  author={Yan, Ke-Xiong and Liu, Yang and Xiao, Yang and Lin, Jun-Hao and Song, Jie and Chen, Ye-Hong and Nori, Franco and Xia, Yan},
  journal={arXiv preprint arXiv:2510.22905},
  year={2025},
  url={https://arxiv.org/abs/2510.22905}
}

@article{7frd-pf1m,
  title = {Cavity QED Based on Strongly Localized Modes: Exponentially Enhancing Single-Atom Cooperativity},
  author = {Bin, Qian and Wu, Ying and Gao, Jin-Hua and Chen, Aixi and Nori, Franco and L\"u, Xin-You},
  journal = {Phys. Rev. Lett.},
  volume = {135},
  issue = {10},
  pages = {103602},
  numpages = {7},
  year = {2025},
  month = {Sep},
  publisher = {American Physical Society},
  doi = {10.1103/7frd-pf1m},
  url = {https://link.aps.org/doi/10.1103/7frd-pf1m}
}

@article{RevModPhys.93.025005,
  title = {Circuit quantum electrodynamics},
  author = {Blais, Alexandre and Grimsmo, Arne L. and Girvin, S. M. and Wallraff, Andreas},
  journal = {Rev. Mod. Phys.},
  volume = {93},
  issue = {2},
  pages = {025005},
  numpages = {72},
  year = {2021},
  month = {May},
  publisher = {American Physical Society},
  doi = {10.1103/RevModPhys.93.025005},
  url = {https://link.aps.org/doi/10.1103/RevModPhys.93.025005}
}

@article{GU20171,
title = {Microwave photonics with superconducting quantum circuits},
journal = {Physics Reports},
volume = {718-719},
pages = {1-102},
year = {2017},
issn = {0370-1573},
doi = {https://doi.org/10.1016/j.physrep.2017.10.002},
url = {https://www.sciencedirect.com/science/article/pii/S0370157317303290},
author = {Xiu Gu and Anton Frisk Kockum and Adam Miranowicz and Yu-xi Liu and Franco Nori}
}

@article{haroche2020cavity,
  title={From cavity to circuit quantum electrodynamics},
  author={Haroche, S and Brune, M and Raimond, JM},
  journal={Nature Physics},
  volume={16},
  number={3},
  pages={243--246},
  year={2020},
  publisher={Nature Publishing Group UK London},
  url={https://www.nature.com/articles/s41567-020-0812-1}
}

@article{PhysRevA.97.062318,
  title = {Quantum memory and gates using a $\mathrm{\ensuremath{\Lambda}}$-type quantum emitter coupled to a chiral waveguide},
  author = {Li, Tao and Miranowicz, Adam and Hu, Xuedong and Xia, Keyu and Nori, Franco},
  journal = {Phys. Rev. A},
  volume = {97},
  issue = {6},
  pages = {062318},
  numpages = {11},
  year = {2018},
  month = {Jun},
  publisher = {American Physical Society},
  doi = {10.1103/PhysRevA.97.062318},
  url = {https://link.aps.org/doi/10.1103/PhysRevA.97.062318}
}

@article{PhysRevLett.117.240501,
  title = {Quantum Networks with Chiral-Light--Matter Interaction in Waveguides},
  author = {Mahmoodian, Sahand and Lodahl, Peter and S\o{}rensen, Anders S.},
  journal = {Phys. Rev. Lett.},
  volume = {117},
  issue = {24},
  pages = {240501},
  numpages = {6},
  year = {2016},
  month = {Dec},
  publisher = {American Physical Society},
  doi = {10.1103/PhysRevLett.117.240501},
  url = {https://link.aps.org/doi/10.1103/PhysRevLett.117.240501}
}

@article{doi:10.1126/sciadv.abb8780,
author = {B. Kannan  and D. L. Campbell  and F. Vasconcelos  and R. Winik  and D. K. Kim  and M. Kjaergaard  and P. Krantz  and A. Melville  and B. M. Niedzielski  and J. L. Yoder  and T. P. Orlando  and S. Gustavsson  and W. D. Oliver },
title = {Generating spatially entangled itinerant photons with waveguide quantum electrodynamics},
journal = {Science Advances},
volume = {6},
number = {41},
pages = {eabb8780},
year = {2020},
doi = {10.1126/sciadv.abb8780},
URL = {https://www.science.org/doi/abs/10.1126/sciadv.abb8780}}

@article{lin2022deterministic,
  title={Deterministic loading of microwaves onto an artificial atom using a time-reversed waveform},
  author={Lin, Wei-Ju and Lu, Yong and Wen, Ping Yi and Cheng, Yu-Ting and Lee, Ching-Ping and Lin, Kuan Ting and Chiang, Kuan Hsun and Hsieh, Ming Che and Chen, Ching-Yeh and Chien, Chin-Hsun and others},
  journal={Nano Letters},
  volume={22},
  number={20},
  pages={8137--8142},
  year={2022},
  publisher={ACS Publications},
  url={https://pubs.acs.org/doi/10.1021/acs.nanolett.2c02578}
}

@article{PhysRevApplied.17.054021,
  title = {Multinode State Transfer and Nonlocal State Preparation via a Unidirectional Quantum Network},
  author = {Ai, Hao and Fang, Ying-Y\"u and Feng, Cheng-Rui and Peng, Zhihui and Xiang, Ze-Liang},
  journal = {Phys. Rev. Appl.},
  volume = {17},
  issue = {5},
  pages = {054021},
  numpages = {11},
  year = {2022},
  month = {May},
  publisher = {American Physical Society},
  doi = {10.1103/PhysRevApplied.17.054021},
  url = {https://link.aps.org/doi/10.1103/PhysRevApplied.17.054021}
}

@article{PhysRevX.7.011035,
  title = {Intracity Quantum Communication via Thermal Microwave Networks},
  author = {Xiang, Ze-Liang and Zhang, Mengzhen and Jiang, Liang and Rabl, Peter},
  journal = {Phys. Rev. X},
  volume = {7},
  issue = {1},
  pages = {011035},
  numpages = {11},
  year = {2017},
  month = {Mar},
  publisher = {American Physical Society},
  doi = {10.1103/PhysRevX.7.011035},
  url = {https://link.aps.org/doi/10.1103/PhysRevX.7.011035}
}

@article{PhysRevLett.128.203602,
  title = {Nonreciprocal Single-Photon Band Structure},
  author = {Tang, Jiang-Shan and Nie, Wei and Tang, Lei and Chen, Mingyuan and Su, Xin and Lu, Yanqing and Nori, Franco and Xia, Keyu},
  journal = {Phys. Rev. Lett.},
  volume = {128},
  issue = {20},
  pages = {203602},
  numpages = {8},
  year = {2022},
  month = {May},
  publisher = {American Physical Society},
  doi = {10.1103/PhysRevLett.128.203602},
  url = {https://link.aps.org/doi/10.1103/PhysRevLett.128.203602}
}

@article{PhysRevX.13.021039,
  title = {Resonance Fluorescence of a Chiral Artificial Atom},
  author = {Joshi, Chaitali and Yang, Frank and Mirhosseini, Mohammad},
  journal = {Phys. Rev. X},
  volume = {13},
  issue = {2},
  pages = {021039},
  numpages = {27},
  year = {2023},
  month = {Jun},
  publisher = {American Physical Society},
  doi = {10.1103/PhysRevX.13.021039},
  url = {https://link.aps.org/doi/10.1103/PhysRevX.13.021039}
}

@article{PhysRevLett.121.043602,
  title = {Observation of the Unconventional Photon Blockade in the Microwave Domain},
  author = {Vaneph, Cyril and Morvan, Alexis and Aiello, Gianluca and F\'echant, Mathieu and Aprili, Marco and Gabelli, Julien and Est\`eve, J\'er\^ome},
  journal = {Phys. Rev. Lett.},
  volume = {121},
  issue = {4},
  pages = {043602},
  numpages = {5},
  year = {2018},
  month = {Jul},
  publisher = {American Physical Society},
  doi = {10.1103/PhysRevLett.121.043602},
  url = {https://link.aps.org/doi/10.1103/PhysRevLett.121.043602}
}

@article{PhysRevLett.121.043601,
  title = {Observation of the Unconventional Photon Blockade},
  author = {Snijders, H. J. and Frey, J. A. and Norman, J. and Flayac, H. and Savona, V. and Gossard, A. C. and Bowers, J. E. and van Exter, M. P. and Bouwmeester, D. and L\"offler, W.},
  journal = {Phys. Rev. Lett.},
  volume = {121},
  issue = {4},
  pages = {043601},
  numpages = {5},
  year = {2018},
  month = {Jul},
  publisher = {American Physical Society},
  doi = {10.1103/PhysRevLett.121.043601},
  url = {https://link.aps.org/doi/10.1103/PhysRevLett.121.043601}
}

@article{PhysRevA.106.013714,
  title = {Quantum trajectory theory and simulations of nonlinear spectra and multiphoton effects in waveguide-QED systems with a time-delayed coherent feedback},
  author = {Crowder, Gavin and Ramunno, Lora and Hughes, Stephen},
  journal = {Phys. Rev. A},
  volume = {106},
  issue = {1},
  pages = {013714},
  numpages = {14},
  year = {2022}, 
  month = {Jul},
  publisher = {American Physical Society},
  doi = {10.1103/PhysRevA.106.013714},
  url = {https://link.aps.org/doi/10.1103/PhysRevA.106.013714}
}

@article{PhysRevA.96.053805,
  title = {Dissipation-induced photonic-correlation transition in waveguide-QED systems},
  author = {Chen, Zihao and Zhou, Yao and Shen, Jung-Tsung},
  journal = {Phys. Rev. A},
  volume = {96},
  issue = {5},
  pages = {053805},
  numpages = {10},
  year = {2017},
  month = {Nov},
  publisher = {American Physical Society},
  doi = {10.1103/PhysRevA.96.053805},
  url = {https://link.aps.org/doi/10.1103/PhysRevA.96.053805}
}

@article{chen2026scalable,
  title={Scalable quantum simulator with an extended gate set in giant atoms},
  author={Chen, Guangze and Kockum, Anton Frisk},
  journal={Quantum},
  volume={10},
  pages={1992},
  year={2026},
  publisher={Verein zur F{\"o}rderung des Open Access Publizierens in den Quantenwissenschaften},
  url={https://quantum-journal.org/papers/q-2026-01-30-1992/}
}

@article{PhysRevA.111.043713,
  title = {Targeted quantum routing of single photons in a giant-atom waveguide-QED system},
  author = {Wang, Yong and Li, Wen-An and Chen, Yuan},
  journal = {Phys. Rev. A},
  volume = {111},
  issue = {4},
  pages = {043713},
  numpages = {10},
  year = {2025},
  month = {Apr},
  publisher = {American Physical Society},
  doi = {10.1103/PhysRevA.111.043713},
  url = {https://link.aps.org/doi/10.1103/PhysRevA.111.043713}
}

@article{Wang:21,
author = {Chen Wang and Xiao-San Ma and Mu-Tian Cheng},
journal = {Opt. Express},
number = {24},
pages = {40116--40124},
publisher = {Optica Publishing Group},
title = {Giant atom-mediated single photon routing between two waveguides},
volume = {29},
month = {Nov},
year = {2021},
url = {https://opg.optica.org/oe/abstract.cfm?URI=oe-29-24-40116},
doi = {10.1364/OE.444096},
}

@article{PhysRevA.76.042319,
  title = {Charge-insensitive qubit design derived from the Cooper pair box},
  author = {Koch, Jens and Yu, Terri M. and Gambetta, Jay and Houck, A. A. and Schuster, D. I. and Majer, J. and Blais, Alexandre and Devoret, M. H. and Girvin, S. M. and Schoelkopf, R. J.},
  journal = {Phys. Rev. A},
  volume = {76},
  issue = {4},
  pages = {042319},
  numpages = {19},
  year = {2007},
  month = {Oct},
  publisher = {American Physical Society},
  doi = {10.1103/PhysRevA.76.042319},
  url = {https://link.aps.org/doi/10.1103/PhysRevA.76.042319}
}

@article{you2011atomic,
  title={Atomic physics and quantum optics using superconducting circuits},
  author={You, Jian-Qiang and Nori, Franco},
  journal={Nature},
  volume={474},
  number={7353},
  pages={589--597},
  year={2011},
  publisher={Nature Publishing Group UK London},
  url = {https://www.nature.com/articles/nature10122}
}

@article{PhysRevLett.97.113602,
  title = {Generation of Narrow-Bandwidth Paired Photons: Use of a Single Driving Laser},
  author = {Kolchin, Pavel and Du, Shengwang and Belthangady, Chinmay and Yin, G. Y. and Harris, S. E.},
  journal = {Phys. Rev. Lett.},
  volume = {97},
  issue = {11},
  pages = {113602},
  numpages = {4},
  year = {2006},
  month = {Sep},
  publisher = {American Physical Society},
  doi = {10.1103/PhysRevLett.97.113602},
  url = {https://link.aps.org/doi/10.1103/PhysRevLett.97.113602}
}

\appendix

\section{Input-output Relation }
\label{sec:Input-output Relation}

In this section, we summarize how to derive Eq.~\eqref{eq:rl-input-output}. We start from the dynamic equations of $r$, which can be written as
\begin{equation}
\dot{r}_{\omega}(t)=-i(\omega-\omega_0)r_{\omega}(t) -iG_r^*(\omega)\sigma_{-}(t),
\end{equation}
where $G_r^*(\omega)={g}_{1}^{*}(t)+g_{2}^{*}(t)e^{-i\varphi_0}e^{-i(\omega-\omega_{0})\tau}$. We choose the initial time $t_0$ or the final time $t_f$ as the initial conditions. Its solutions are
\begin{equation}
\label{eq:input output equation}
\begin{aligned}
r_{\omega}(t)=&e^{-i(\omega-\omega_0)(t-t_0)}r_{\omega}(t_{0})\\
&-i\int_{t_{0}}^{t}dt^{\prime}e^{-i(\omega-\omega_0)(t-t^{\prime})}G_r^{*}(w)\sigma_{-}(t^{\prime}),\\
r_{\omega}(t)=&e^{-i(\omega-\omega_0)(t-t_f)}r_{\omega}(t_{f})\\
&+i\int_{t}^{t_f}dt^{\prime}e^{-i(\omega-\omega_0)(t-t^{\prime})}G_r^{*}(w)\sigma_{-}(t^{\prime}).
\end{aligned}
\end{equation}
The input and output operators are defined as
\begin{equation}
\label{eq:input output definition}
\begin{aligned}
r_{in}(t)=\frac{1}{\sqrt{2\pi}}\int d\omega\cdot e^{-i(\omega-\omega_0)(t-t_{0})}r_{\omega}(t_{0}),\\
r_{out}(t)=\frac{1}{\sqrt{2\pi}}\int d\omega\cdot e^{-i(\omega-\omega_0)(t-t_{f})}r_{\omega}(t_{f}).
\end{aligned}
\end{equation}
Integrating Eq.~\eqref{eq:input output equation} over frequency and combining it with Eq.~\eqref{eq:input output definition}, we obtain
\begin{equation}
\label{eq:input output omega}
\begin{aligned}
r_{out}(t)=&r_{in}(t)-\frac{1}{\sqrt{2\pi}}\int d\omega\{
i\int_{t_{0}}^{t_f}dt^{\prime}e^{-i(\omega-\omega_0)(t-t^{\prime})}
\\&[{g}_{1}^{*}(t)+g_{2}^{*}(t)e^{-i\varphi_0}e^{-i(\omega-\omega_{0})\tau}]\sigma_{-}(t^{\prime})\}.
\end{aligned}
\end{equation}
After integration, we arrive at the input-output relation of $r$,  
\begin{equation}r_{\mathrm{out}}(t)
= r_{\mathrm{in}}(t)
  - i\sqrt{2\pi}g_1^{*}\sigma_{-}(t)
  - i\sqrt{2\pi}g_2^{*}\sigma_{-}(t+\tau)e^{-i\varphi_0}.
  \end{equation}
Following the same derivation of the above equation, the dynamic equations of $\ell$ are
\begin{equation}
\dot{{\ell}}_{\omega}(t)=-i(\omega-\omega_0){\ell}_{\omega}(t) -iG_{\ell}^*(\omega)\sigma_{-}(t),
\end{equation}
where $G_{\ell}^*(\omega)={g}_{1}^{*}(t)+g_{2}^{*}(t)e^{i\varphi_0}e^{i(\omega-\omega_{0})\tau}$. The input-output relation of $\ell$ is
\begin{equation}\ell_{\mathrm{out}}(t)
= \ell_{\mathrm{in}}(t)
  - i\sqrt{2\pi}g_1^{*}\sigma_{-}(t)
  - i\sqrt{2\pi}g_2^{*}\sigma_{-}(t-\tau)e^{i\varphi_0}.
\end{equation}
The distance between the two coupling points of the giant atom brings two main differences. One lies in the phase ($e^{i\varphi_0}$ and $e^{-i\varphi_0}$), while the other is in $\tau$. While the pulse reaches the first coupling point, the system time is $t$. When it reaches the second coupling point, the system time is $t+\tau$ or $t -\tau$. As we denote the first coupling point located in $x=0$, the effect of $\tau$ appears only in the last term, which represents the contribution of the second coupling point. It might be confusing that the term involving $t+\tau$ violates causality, while it is a result of an additional phase term in Eq.~\eqref{eq:input output omega}, i.e., $e^{-i(\omega-\omega_{0})\tau}$. Similar results have also been reported in Ref.~\cite{PhysRevA.88.043806}.

In this article, we focus on the regime that $\tau$ to be much smaller than the atom lifetime~\cite{ask2022non}, where $\tau$ is negligible. After neglecting $\tau$, the final input-output relations are
\begin{equation}
{\label{eq:input-output}}
\begin{aligned}
r_{\mathrm{out}}(t)
&= r_{\mathrm{in}}(t)
  - i\sqrt{2\pi}g_1^{*}\sigma_{-}(t)
  - i\sqrt{2\pi}g_2^{*}\sigma_{-}(t)e^{-i\varphi_0},
  \\
\ell_{\mathrm{out}}(t)
&= \ell_{\mathrm{in}}(t)
  - i\sqrt{2\pi}g_1^{*}\sigma_{-}(t)
  - i\sqrt{2\pi}g_2^{*}\sigma_{-}(t)e^{i\varphi_0},
  \end{aligned}
  \end{equation}
which are Eq.~\eqref{eq:rl-input-output} in the main text.


\section{Derivation of the Langevin Equation }
\label{sec:Derivation of the Langevin Equation}

In this section, we show how to derive Eq.~\eqref{eq:X equation}. We start from the Heisenberg equation for the $\sigma_-$,
\begin{equation}
    \dot{\sigma}_-=-i[\sigma_-,H].
\end{equation}

By solving the field operators and replacing them, we obtain
\begin{equation}
\begin{aligned}
\dot{\sigma}_{-}(t)=&-i[{\sigma}_{-}(t),\frac{1}{2}\Delta \sigma_z(t)]-[\sigma_{-}(t),\sigma_{+}(t)] 
\int_{-\infty}^{\infty} d\omega 
\\&\Big\{G_r(\omega)e^{-i(\omega-\omega_0)(t-t_0)}r_{\omega}(t_{0})
\\&+G_{\ell}(\omega)e^{-i(\omega-\omega_0)(t-t_{0})}l_{\omega}(t_{0}) \\
& -i\int_{t_{0}}^{t}dt^{\prime}e^{-i(\omega-\omega_0)(t-t^{\prime})}|G_r(\omega)|^2\sigma_{-}(t^{\prime})
\\&-i\int_{t_{0}}^{t}dt^{\prime}e^{-i(\omega-\omega_0)(t-t^{\prime})}|G_{\ell}(\omega)|^2\sigma_{-}(t^{\prime})
\Big\}.
\end{aligned}
\end{equation}
The single-integral term is directly obtained by expanding the coefficients $G_{\ell}$ and $G_{r}$, 
\begin{equation}
\begin{aligned}
&\int_{-\infty}^{\infty}d\omega\{ G_r(\omega)e^{-i(\omega-\omega_0)(t-t_0)}r_{\omega}^{\dagger}(t_{0})\\& \qquad+G_{\ell}(\omega)e^{-i(\omega-\omega_0)(t-t_{0})}{\ell}^{\dagger}_\omega(t_{0})\}
\\
=&\int_{-\infty}^{\infty}d\omega\{[{g}_{1}+g_{2}e^{i\varphi_0}e^{i(\omega-\omega_{0})\tau}] \\&\times e^{-i(\omega-\omega_0)(t-t_0)}r_{\omega}^{\dagger}(t_{0})+[{g}_{1}+g_{2}e^{-i\varphi_0}e^{-i(\omega-\omega_{0})\tau}]
\\&\times
e^{-i(\omega-\omega_0)(t-t_{0})}\ell^{\dagger}_\omega(t_{0})]\}.
\end{aligned}
\end{equation}
Integrating four terms individually, the results are
\begin{equation}
{\label{eq:1st integral}}
\begin{aligned} 
& \int_{-\infty}^{\infty}d\omega g_{1}e^{-i(\omega-\omega_0)(t-t_0)}r_{\omega}(t_{0})=\sqrt{2\pi}r_{in}(t)g_{1},  
\\ & \int_{-\infty}^{\infty}d\omega g_{1} e^{-i(\omega-\omega_0)(t-t_{0})}\ell_{\omega}(t_{0})=\sqrt{2\pi}\ell_{in}(t)g_{1}  ,
\\&
\int_{-\infty}^\infty d\omega  g_2 e^{i\varphi_0} e^{-i(\omega-\omega_0)(t-t_0-\tau)} r_\omega(t_0)
\\ &\qquad = \sqrt{2\pi} r_\mathrm{in}(t-\tau) g_2 e^{i\varphi_0},
\\ 
&
\int_{-\infty}^\infty d\omega g_2 e^{-i\varphi_0} e^{-i(\omega-\omega_0)(t-t_0+\tau)} \ell_\omega(t_0)
\\&\qquad= \sqrt{2\pi}\ell_\mathrm{in}(t+\tau)g_2 e^{-i\varphi_0}.
\end{aligned}
\end{equation}
While double-integral terms are more involved, we expand the coefficients $|G_{\ell}|^2$ and $|G_{r}|^2$, resulting in
\begin{equation}
\begin{aligned}
|G_r(\omega)|^2=&|g_{1}|^{2}+|g_{2}|^{2}+g_{1} g_{2}^{*} e^{-i\varphi_0}e^{-i(\omega-\omega_{0})\tau} \\&
+g_{1}^{*}g_{2}e^{i\varphi_0}e^{i(\omega-\omega_{0})\tau},
\\
|G_\ell(\omega)|^2=&|g_{1}|^{2}+|g_{2}|^{2}+ g_{1}g_{2}^{*}e^{i\varphi_0}e^{i(\omega-\omega_{0})\tau}\\&
+g_{1}^{*} g_{2} e^{-i\varphi_0}e^{-i(\omega-\omega_{0})\tau}.
\end{aligned}
\end{equation}
Integrating the first two terms of $|G_{\ell}|^2$ and $|G_{r}|^2$ is simple, as they do not involve time or phase contributions. Integration is given by
\begin{equation}
\begin{aligned}
&i\int_{t_{0}}^{t}dt^{\prime}\int_{-\infty}^{\infty}d\omega\{2[|g_{1}|^{2}+|g_{2}|^{2}] e^{-i\omega(t-t^{\prime})}\sigma_{-}(t^{\prime})\\&=2i\int_{t_{0}}^{t}dt^{\prime}\{[|g_{1}|^{2}+|g_{2}|^{2}] 2\pi\delta(t-t')\sigma_{-}(t^{\prime})\\&=2\pi i [|g_{1}|^{2}+|g_{2}|^{2}] \sigma_{-}(t).
\end{aligned}
\end{equation}
Integrating the last two terms of $|G_{\ell}|^2$ and $|G_{r}|^2$ gives
\begin{equation}
\begin{aligned}
=& i\int_{-\infty}^{\infty}d\omega\int_{t_{0}}^{t}dt^{\prime}\{e^{-i(\omega-\omega_{0})(t-t^{\prime})}[g_{1} g_{2}^{*} e^{-i\phi_0}e^{-i(\omega-\omega_{0})\tau} 
\\&
+g_{1}^{*}g_{2}e^{i\phi_0}e^{i(\omega-\omega_{0})\tau}]+e^{-i(\omega-\omega_{0})(t-t^{\prime})} \\&\times[ g_{1}g_{2}^{*}e^{i\varphi_0}e^{i(\omega-\omega_{0})\tau}
+g_{1}^{*}  g_{2} e^{-i\varphi_0}e^{-i(\omega-\omega_{0})\tau}
]\sigma_{-}(t^{\prime})\}
\\= & i\int_{-\infty}^{\infty}d\omega \int_{t_0}^{t}dt' 
 \{[g_{1}g_{2}^{*}e^{-i\phi_{0}}e^{-i\omega(t-t^{\prime}+\tau)}
 \\&+g_{1}^{*}g_{2}e^{i\phi_{0}}e^{-i(\omega-\omega_{0})(t-t^{\prime}-\tau)}\\ &+
g_{1}g_{2}^{*}e^{i\phi_{0}}e^{-i(\omega-\omega_{0})(t-t^{\prime}-\tau)}\\ &
+g_{1}^{*}g_{2}e^{-i\phi_{0}}e^{-i(\omega-\omega_{0})(t-t^{\prime}+\tau)}
]\sigma_{-}(t^{\prime})\}.
\end{aligned}
\end{equation}
Integrating frequency first, then time, we have
\begin{equation}
{\label{eq:2nd integral}}
\begin{aligned}=&
 2\pi i\int_{t_0}^{t} dt'\Big\{\big[g_{1}g_{2}^{*}e^{-i\phi_{0}}\delta(t-t'+\tau)\\&+ g_{1}^{*}g_{2}e^{i\phi_{0}}\delta(t-t'-\tau) + g_{1}g_{2}^{*}e^{i\phi_{0}}\delta(t-t'-\tau)\\&+ g_{1}^{*}g_{2}e^{-i\phi_{0}}\delta(t-t'+\tau)\big]\sigma_{-}(t')\Big\}
\\
=&4\pi iRe(g_{1}g_{2}^{*})e^{i\phi_{0}}\sigma_{-}(t-\tau).
\end{aligned}
\end{equation}
There are only two terms left, while the other two terms vanish since $t'=t+\tau$ does not lie in the integration region.

Combining Eq.~(\ref{eq:1st integral}), (\ref{eq:2nd integral}), and the commutation relation of the Pauli matrix, 
\begin{equation}[{\sigma}_{-},{\sigma}_{+}] =-\sigma_z, [{\sigma}_{-},\sigma_z]=2\sigma_-,\end{equation}
we have the Langevin equation of $\sigma_-$
\begin{equation}
\begin{aligned}
&\dot{\sigma}_{-}(t)=-i\Delta\sigma_-(t) +i\sigma_z(t)\\&\times\Big\{\sqrt{2\pi}r_{\mathrm{in}}(t)g_{1}+\sqrt{2\pi}l_{\mathrm{in}}(t)g_{1}\\ &+\sqrt{2\pi}r_{\mathrm{in}}(t-\tau)g_{2}e^{i\varphi_{0}}+\sqrt{2\pi}\ell_{\mathrm{in}}(t+\tau)g_{2}e^{-i\varphi_{0}}
\\ &-2\pi i [(|g_{1}|^{2}+|g_{2}|^{2}) \sigma_{-}(t)+2\mathrm{Re}(g_{1}g_{2}^{*})e^{i\phi_{0}}\sigma_{-}(t-\tau)]\Big\}.
\end{aligned}
\end{equation}
The first term represents the contribution induced by the atom, i.e., free evolution. The second line arises from the incident field at the first coupling point, and the third line originates from the time-delayed input at the second coupling point. The fourth line corresponds to the backaction term resulting from the atom–photon interaction, which retains a time-delayed contribution here. 

As discussed in Appendix~\ref{sec:Input-output Relation}, we still ignore $\tau$. The Langevin equation of $\sigma_-$ is simplified into
\begin{equation}
\label{eq:Langevin equation}
\begin{aligned}
\dot{\sigma}_-(t)
=& -{i}\Delta\sigma_-(t)
+i\sigma_z(t)\Big\{
\sqrt{2\pi}r_{\mathrm{in}}(t)g_1
+\sqrt{2\pi}l_{\mathrm{in}}(t)g_1
\\&
+\sqrt{2\pi}r_{\mathrm{in}}(t)g_2e^{i\varphi_{0}}
+\sqrt{2\pi}\ell_{\mathrm{in}}(t)g_2e^{-i\varphi_{0}}
\\
&
-2\pi i[\big(|g_1|^2+|g_2|^2\big)\sigma_{-}(t)
+2\mathrm{Re}(g_1 g_2^{*})e^{i\varphi_{0}}\sigma_{-}(t)
]\Big\},
\end{aligned}
\end{equation}

The equation of $\langle0|\sigma_{-}(t) r_{\mathrm{in}}^{\dagger}(k)|0\rangle$ is obtained by sandwiching Eq.~\eqref{eq:Langevin equation} between $\langle0|$ and $r_{\mathrm{in}}^{\dagger}(k)|0\rangle$. We use 
\begin{equation}
\label{eq:lr product}
    \begin{aligned}
    \langle0|\sigma_z(t)\sigma_{-}(t)r_{\mathrm{in}}^{\dagger}(k)|0\rangle&=-\langle0|\sigma_{-}(t)r_{\mathrm{in}}^{\dagger}(k)|0\rangle,
    \\
    \langle0|\sigma_z(t)r_{\mathrm{in}}(t)r_{\mathrm{in}}^{\dagger}(k)|0\rangle&=-\langle0|r_{\mathrm{in}}(t)r_{\mathrm{in}}^{\dagger}(k)|0\rangle,
    \\
    \langle0|\sigma_z(t)\ell_{\mathrm{in}}(t)r_{\mathrm{in}}^{\dagger}(k)|0\rangle&=-\langle0|\ell_{\mathrm{in}}(t)r_{\mathrm{in}}^{\dagger}(k)|0\rangle=0,
    \end{aligned}
\end{equation}
since the atom is in the ground state $|0\rangle$, and $[\ell_k,r_k]=0$. Denoting $X(t)=\langle0|\sigma_-(t)r_{\mathrm{in}}^{\dagger}(k)|0\rangle$, we obtain 
\begin{equation}
\begin{aligned}
\dot{X}
=& -{i}\Delta X
-i(g_1+g_2e^{i\varphi_{0}}) e^{-ikt}\\&-
2\pi \big(|g_1|^2+|g_2|^2\big)X
-4\pi \mathrm{Re}(g_1 g_2^{*})e^{i\phi_{0}}X,
\end{aligned}
\end{equation}
which is Eq.~\eqref{eq:X equation} in the main text. 


\section{Single-photon Scattering Matrices for Left-propagating Mode  }
\label{sec:Single-photon Scattering Matrices for Left-propagating Mode}

Following the same procedure as in the main text, the single-photon scattering matrices for a photon incident from the left-propagating mode are
\begin{equation}
\begin{aligned}
\langle 0| r_{\mathrm{out}}(p) \ell_{\mathrm{in}}^{\dagger}(k)|0\rangle=\delta(k-p)r_{\ell},
\\
\langle 0| \ell_{\mathrm{out}}(p) \ell_{\mathrm{in}}^{\dagger}(k)|0\rangle=\delta(k-p)t_{\ell},
\end{aligned}
\end{equation}
with
\begin{equation}
\begin{aligned}
r_{\ell}=\frac{-2\pi i[|g_1|^2+|g_2|^2e^{-2i\phi_0}+2\mathrm{Re}(g_1 g_2^{*})e^{-i\phi_0}] }{k-\Delta+2\pi i(|g_1|^2+|g_2|^2)+4\pi i\mathrm{Re}(g_1 g_2^{*})e^{i\phi_{0}}},
\\
t_{\ell}=\frac{k-\Delta-4\pi  g^*_1g_2\sin(\phi_0)}{k-\Delta+2\pi i(|g_1|^2+|g_2|^2)+4\pi i\mathrm{Re}(g_1 g_2^{*})e^{i\phi_{0}}}.
\end{aligned}\end{equation}
One can find $|r_{\ell}|^2+|t_{\ell}|^2=1$, $|r_{\ell}|^2=|r_{r}|^2$, and $|t_{\ell}|^2+|t_{\ell}|^2=1$. Thus, even when $g_1\neq g_2$, the reflection and transmission probabilities are identical for the two incidence directions, which is a consequence of symmetry. 


\section{Additional Derivation of the Two-photon Scattering Matrices }
\label{sec:Additional Derivation of the Two-photon Scattering Matrices}

\subsection{Both Photons are Transmitted}
\label{sec:Both Photons are Transmitted}

We first derive the second term in Eq.~\eqref{eq:rrrr}. Following the same method in Sec.~\ref{sec:Single-photon Scattering Matrices}, we sandwich Eq.~\eqref{eq:Langevin equation} between $\langle0|r_{\mathrm{in}}(p_1)$ and $r_{\mathrm{in}}^{\dagger}(k_1)r_{\mathrm{in}}^{\dagger}(k_2)|0\rangle$, resulting in 
\begin{equation}
\label{eq:2ndsigma-equation}
\begin{aligned}
&\quad\big\langle 0 \big| r_{\mathrm{in}}(p_{1}) \dot{\sigma}_{-}(t)
r_{\mathrm{in}}^{\dagger}(k_{1}) r_{\mathrm{in}}^{\dagger}(k_{2}) \big| 0 \big\rangle
\\
=& - i\Delta 
\big\langle 0 \big| r_{\mathrm{in}}(p_{1}) \sigma_{-}(t)
r_{\mathrm{in}}^{\dagger}(k_{1}) r_{\mathrm{in}}^{\dagger}(k_{2}) \big| 0 \big\rangle
\\[4pt]
&+ \sqrt{2\pi}i (g_1+g_2e^{i\phi_0})
\\&\times\big\langle 0 \big| r_{\mathrm{in}}(p_{1}) \sigma_{z}(t) r_{\mathrm{in}}(t)
r_{\mathrm{in}}^{\dagger}(k_{1}) r_{\mathrm{in}}^{\dagger}(k_{2}) \big| 0 \big\rangle
\\[4pt]
& + \sqrt{2\pi}i (g_1+g_2e^{-i\phi_0})
\\&\times
\big\langle 0 \big| r_{\mathrm{in}}(p_{1}) \sigma_{z}(t) \ell_{\mathrm{in}}(t)
r_{\mathrm{in}}^{\dagger}(k_{1}) r_{\mathrm{in}}^{\dagger}(k_{2}) \big| 0 \big\rangle
\\[4pt]
& +  \!\left[2\pi(|g_{1}|^{2}+|g_{2}|^{2}) + 4\pi\mathrm{Re}\!\left(g_{1}g_{2}^{*}\right)
e^{+i\varphi_{0}}\right]\\
&\times
\big\langle 0 \big| r_{\mathrm{in}}(p_{1}) \sigma_{z}(t)\sigma_{-}(t)
r_{\mathrm{in}}^{\dagger}(k_{1}) r_{\mathrm{in}}^{\dagger}(k_{2}) \big| 0 \big\rangle .
\end{aligned}\end{equation}
The fourth term can be simplified with $\sigma_z\sigma_-=-\sigma_-$, while the third term equals $0$. The scattering matrix of the second term can be written as
\begin{equation}
\label{eq:sigmaz-simplify}
\begin{aligned}
&\big\langle 0 \big| r_{\mathrm{in}}(p_{1}) \sigma_{z}(t) r_{\mathrm{in}}(t)
r_{\mathrm{in}}^{\dagger}(k_{1}) r_{\mathrm{in}}^{\dagger}(k_{2}) \big| 0 \big\rangle
\\=&\big\langle 0 \big| r_{\mathrm{in}}(p_{1}) (2\sigma_{+}(t)\sigma_{-}(t)-1) r_{\mathrm{in}}(t)
r_{\mathrm{in}}^{\dagger}(k_{1})r_{\mathrm{in}}^{\dagger}(k_{2}) \big| 0 \big\rangle
\\=&\big\langle 0 \big| r_{\mathrm{in}}(p_{1})(2\sigma_{+}(t)\sigma_{-}(t)-1)r_{\mathrm{in}}^{\dagger}(k_{2}) \big| 0 \big\rangle\frac{e^{-ik_1t}}{\sqrt{2\pi}}
\\&+\big\langle 0 \big|r_{\mathrm{in}}(p_{1})(2\sigma_{+}(t)\sigma_{-}(t)-1)r_{\mathrm{in}}^{\dagger}(k_{1}) \big| 0 \big\rangle\frac{e^{-ik_2t}}{\sqrt{2\pi}}
\\=&[2\big\langle 0 \big| r_{\mathrm{in}}(p_{1})\sigma_{+}(t)\big| 0 \big\rangle\big\langle 0\big| \sigma_{-}(t)r_{\mathrm{in}}^{\dagger}(k_{2}) \big| 0 \big\rangle
\\&-\big\langle 0 \big| r_{\mathrm{in}}(p_{1})r_{\mathrm{in}}^{\dagger}(k_{2}) \big| 0 \big\rangle]\frac{e^{-ik_1t}}{\sqrt{2\pi}}
\\&+[2\big\langle 0 \big| r_{\mathrm{in}}(p_{1})\sigma_{+}(t)\big| 0 \big\rangle\big\langle 0\big| \sigma_{-}(t)r_{\mathrm{in}}^{\dagger}(k_{2}) \big| 0 \big\rangle
\\&-\big\langle 0 \big| r_{\mathrm{in}}(p_{1})r_{\mathrm{in}}^{\dagger}(k_{1}) \big| 0 \big\rangle]\frac{e^{-ik_2t}}{\sqrt{2\pi}}
\\=& \left[ \frac{1}{\pi} s_r({k_1})s_r^*({p_1}) e^{-i(k_1-p_1)t}
      - \delta(p_1-k_1) \right]\frac{1}{\sqrt{2\pi}}e^{-ik_2 t}
  \\&+ \left[ \frac{1}{\pi} s_r({k_2})s_r^*({p_1}) e^{-i(k_2-p_1)t}
      - \delta(p_1-k_2) \right]\frac{1}{\sqrt{2\pi}}e^{-ik_1 t}.
\end{aligned}\end{equation}
The first equality is the result of $ \sigma_{z}(t) =2\sigma_{+}(t)\sigma_{-}(t)-1$. The second equality can be obtained via Eq.~(\ref{eq:tk-relation}). Note that $r_{\mathrm{in}}(t)$ can be contracted with either $r_{\mathrm{in}}^{\dagger}(k_{1})$ or $r_{\mathrm{in}}^{\dagger}(k_{2})$, which yields two contributions. The third equality is obtained by inserting a complete basis between $ \sigma_{+}(t)$ and $ \sigma_{-}(t)$, while only $ \big| 0 \big\rangle\big\langle 0\big|$ remains. The last equality is obtained through the Fourier transformation of Eq.~\eqref{eq:sr delta}.

With Eq.~\eqref{eq:2ndsigma-equation} and Eq.~\eqref{eq:sigmaz-simplify} and denoting $Y_{r\sigma}=\big\langle 0 \big| r_{\mathrm{in}}(p_{1}) \sigma_{-}(t)
r_{\mathrm{in}}^{\dagger}(k_{1}) r_{\mathrm{in}}^{\dagger}(k_{2}) \big| 0 \big\rangle$, the equation of $Y_{r\sigma}$ is given by 
\begin{equation}
\begin{aligned}
\dot{Y}_{r\sigma}=& - i\Delta Y_{r\sigma}+ i(g_1+g_2e^{i\phi_0})\Big\{\frac{1}{\pi}[s_r({k_1})+s_r({k_2})]s_r^*({p_1})
\\& \times
 e^{-i(k_1+k_2-p_1)t}- \delta(p_{1}-k_{1})e^{-ik_{2}t}- \delta(p_{1}-k_{2})\times \\&e^{-ik_{1}t}\Big\}
- \!\left[2\pi(|g_{1}|^{2}+|g_{2}|^{2})+ 4\pi\mathrm{Re}\!\left(g_{1}g_{2}^{*}\right)e^{i\varphi_{0}}\right] Y_{r\sigma}.
\end{aligned}
\end{equation}

The solution is
\begin{equation}\begin{aligned}
Y_{r\sigma}=&(g_1+g_2e^{i\phi_0})\bigg[-\frac{1}{\pi}\frac{[s_r({k_1})+s_r({k_2})]s_r^*({p_1}) }{{(k_1+k_2-p_1)-\Lambda}}
\\&\times e^{-i(k_1+k_2-p_1)t}+\frac{\delta(p_{1}-k_{1})}{{k_2-\Lambda}}e^{-ik_{2}t}
\\&+\frac{\delta(p_{1}-k_{2})}{{k_1-\Lambda}}e^{-ik_{1}t}\bigg],
\end{aligned}
\end{equation}
where $\Lambda=\Delta-i\left[2\pi(|g_{1}|^{2}+|g_{2}|^{2})+ 4\pi\mathrm{Re}\!\left(g_{1}g_{2}^{*}\right)e^{i\varphi_{0}}\right]$. Using Eq.~(\ref{eq:sr}) to simplify the result and performing a Fourier transform, we obtain
\begin{equation}
\label{eq:rsrr result}
\begin{aligned}
&\big\langle 0 \big| r_{\mathrm{in}}(p_{1}) \sigma_{-}(p_2)
r_{\mathrm{in}}^{\dagger}(k_{1}) r_{\mathrm{in}}^{\dagger}(k_{2}) \big| 0 \big\rangle\\
=&-\frac{1}{\pi}\delta(k_{1}+k_{2}-p_{1}-p_{2})s_r({p_2})s_r^*({p_1}) [s_r({k_1})+s_r({k_2})]
\\&+ s_r({k_1})\delta(k_{2}-p_{1})\delta(k_{1}-p_{2})+s_r({k_2})\delta(k_{1}-p_{1})\delta(k_{2}-p_{2}),
\end{aligned}
\end{equation}
which is Eq.~\eqref{eq:rsrr} in the main text.

The fourth term in Eq.~\eqref{eq:2ndsigma-equation} can be treated in the same way, where the differential equation is written as
\begin{equation}
\label{eq:2ndsigma-equation-l}
\begin{aligned}
&\big\langle 0 \big| \ell_{\mathrm{in}}(p_{1}) \dot{\sigma}_{-}(t)
r_{\mathrm{in}}^{\dagger}(k_{1}) r_{\mathrm{in}}^{\dagger}(k_{2}) \big| 0 \big\rangle
\\
=& - i\Delta 
\big\langle 0 \big| \ell_{\mathrm{in}}(p_{1}) \sigma_{-}(t)
r_{\mathrm{in}}^{\dagger}(k_{1}) r_{\mathrm{in}}^{\dagger}(k_{2}) \big| 0 \big\rangle
\\[4pt]
& + \sqrt{2\pi}i (g_1+g_2e^{i\phi_0}) 
\\[4pt]&\times\big\langle 0 \big| \ell_{\mathrm{in}}(p_{1}) \sigma_{z}(t) r_{\mathrm{in}}(t)
r_{\mathrm{in}}^{\dagger}(k_{1}) r_{\mathrm{in}}^{\dagger}(k_{2}) \big| 0 \big\rangle
\\[4pt]
&+ \sqrt{2\pi}i (g_1+g_2e^{-i\phi_0})
\\[4pt]&\times
\big\langle 0 \big| \ell_{\mathrm{in}}(p_{1}) \sigma_{z}(t) \ell_{\mathrm{in}}(t)
r_{\mathrm{in}}^{\dagger}(k_{1}) r_{\mathrm{in}}^{\dagger}(k_{2}) \big| 0 \big\rangle
\\[4pt]
& +  \!\left[2\pi(|g_{1}|^{2}+|g_{2}|^{2}) + 4\pi\mathrm{Re}\!\left(g_{1}g_{2}^{*}\right)
e^{i\varphi_{0}}\right]
\\[4pt]
& \times\big\langle 0 \big| \ell_{\mathrm{in}}(p_{1}) \sigma_{z}(t)\sigma_{-}(t)
r_{\mathrm{in}}^{\dagger}(k_{1}) r_{\mathrm{in}}^{\dagger}(k_{2}) \big| 0 \big\rangle .
\end{aligned}
\end{equation}
The scattering matrix of the second term is almost identical to Eq.~\eqref{eq:sigmaz-simplify}, except that the photon is reflected. Following the same procedure, we obtain
\begin{equation}
\label{eq:sigmaz-simplify-l}
\begin{aligned}
&\big\langle 0 \big| \ell_{\mathrm{in}}(p_{1}) \sigma_{z} r_{\mathrm{in}}(t)
r_{\mathrm{in}}^{\dagger}(k_{1}) r_{\mathrm{in}}^{\dagger}(k_{2}) \big| 0 \big\rangle
\\=& \left[ \frac{1}{\pi} s_r({k_1})s^*_{\ell}(p_1) e^{-i(k_1-p_1)t}
       \right]\frac{1}{\sqrt{2\pi}}e^{-ik_2 t}
\\&  + \left[ \frac{1}{\pi} s_r({k_2})s^*_{\ell}(p_1)e^{-i(k_2-p_1)t}
      \right]\frac{1}{\sqrt{2\pi}}e^{-ik_1 t} ,
\end{aligned}
\end{equation}
where
\begin{equation}
\label{eq:sl}
\begin{aligned}
&\langle0|\sigma_-(p)\ell_{\mathrm{in}}^{\dagger}(k)|0\rangle=\delta(p-k)s_{\ell}(k),\\
s_{\ell}(k)=&\frac{\sqrt{2\pi}(g_1+g_2e^{-i\phi_0})}{k-\Delta+2\pi i(|g_1|^2+|g_2|^2\big)+4\pi i\mathrm{Re}(g_1 g_2^{*})e^{i\phi_{0}}},
\end{aligned}
\end{equation}
which relates to the scattering amplitude for a single left-propagating photon being absorbed by the atom. 

With Eq.~\eqref{eq:sigmaz-simplify-l}, Eq.~\eqref{eq:2ndsigma-equation-l} can be simplified into 
\begin{equation}
\begin{aligned}
\dot{Y}_{\ell\sigma}=& - i\Delta Y_{\ell\sigma}+ i (g_1+g_2e^{-i\phi_0})\Bigg[\frac{1}{\pi}[s_r({k_1})+s_r({k_2})]\\
&\times s_{\ell}^{*}{(p_1)}e^{-i(k_1+k_2-p_1)t}\Bigg]\\
&- \!\left[2\pi(|g_{1}|^{2}+|g_{2}|^{2})+ 4\pi\mathrm{Re}\!\left(g_{1}g_{2}^{*}\right)e^{i\varphi_{0}}\right] Y_{\ell\sigma} .
\end{aligned}
\end{equation}
Solving this differential equation, we obtain
\begin{equation}
\begin{aligned}
Y_{\ell\sigma}=&-\frac{1}{\pi\sqrt{2\pi}}[s_r({k_1})+s_r({k_2})]\\
&\times s_{\ell}^{*}{(p_1)}s_r({k_1+k_2-p_1})e^{-i(k_1+k_2-p_1)t}.
\end{aligned}
\end{equation}
After the Fourier transformation, we have 
\begin{equation}
\label{eq:lsrr result}
\begin{aligned}
&\big\langle 0 \big| \ell_{\mathrm{in}}(p_{1}) \sigma_{-}(p_2)
r_{\mathrm{in}}^{\dagger}(k_{1}) r_{\mathrm{in}}^{\dagger}(k_{2}) \big| 0 \big\rangle\\
=&-\frac{1}{\pi}\delta(k_{1}+k_{2}-p_{1}-p_{2})s_r({p_{2}})s^{*}_{\ell}({p_{1}})[s_r({k_1})+s_r({k_2})],
\end{aligned}
\end{equation}
which is Eq.~\eqref{eq:lsrr} in the main text.

Using Eq.~\eqref{eq:rsrr result} and Eq.~\eqref{eq:lsrr result}, Eq.~\eqref{eq:rlrr} can be written as
\begin{equation}
\label{eq:rrrr unsimplified}
\begin{aligned}
&\langle0|r_{\mathrm{out}}(p_1)r_{\mathrm{out}}(p_2)r_{\mathrm{in}}^{\dagger}(k_1)r_{\mathrm{in}}^{\dagger}(k_2)|0\rangle
\\=&-t_r({p_1})[\delta(k_{2}-p_{1})\delta(k_{1}-p_{2})+\delta(k_{1}-p_{1})\delta(k_{2}-p_{2})]
\\&-t_r({p_1})i\sqrt{2\pi}\big(g_2^{*}e^{-i\phi_0}+g_1^{*}\big)
\\&\times\{-\frac{1}{\pi}\delta(k_{1}+k_{2}-p_{1}-p_{2})s_r({p_{2}})s^{*}_{r}({p_{1}})[s_r({k_1})+s_r({k_2})]
\\&+ s_r({k_1})\delta(k_{2}-p_{1})\delta(k_{1}-p_{2})+s_r({k_2})\delta(k_{1}-p_{1})\delta(k_{2}-p_{2})\}
\\&-r_{\ell}(p_1)i\sqrt{2\pi}\big(g_2^{*}e^{-i\phi_0}+g_1^{*}\big)
\\&\times\{-\frac{1}{\pi}\delta(k_{1}+k_{2}-p_{1}-p_{2})s_r({p_{2}})s^{*}_{\ell}({p_{1}})[s_r({k_1})+s_r({k_2})]\}.
\end{aligned}
\end{equation}
The terms that correspond to the single-photon process can be simplified by 
\begin{equation}
\label{eq:s to t}
t_r(p_1)-t_r(p_1)\sqrt{2\pi} i(g_1^*+g_2^*e^{-i\phi_0})s_r(k)=t_r(p_1)t_r(k),
\end{equation}
which is the result of Eq.~\eqref{eq:transmission solving} and~\eqref{eq:sr delta}.

The photon-photon interaction term can be simplified by
\begin{equation}
\label{eq:tsrs to s}
\begin{aligned}
&t_r({p_1})s^{*}_r({p_{1}})+r_{\ell}({p_1})s^{*}_{\ell}({p_{1}})
=s_r({p_{1}})\frac{g^*_1+g^*_2e^{-i\phi_0}}{g_1+g_2e^{i\phi_0}},
\end{aligned}
\end{equation}
which is obtained by substituting all scattering amplitudes into the expression. The numerator on the left side of Eq.~\eqref{eq:tsrs to s} can be written as 
\begin{equation}
\begin{aligned}
&{(g^*_1+g^*_2e^{-i\phi_0}})\Big\{\Big[k-\Delta+2\pi ig_1g_2^*\big(e^{i\varphi_{0}}-e^{-i\varphi_{0}}\big)\Big]\\
&\times\Big(g_{1}^{*}+g_{2}^{*}e^{-i\varphi_{0}}\Big)-2\pi i\Big[|g_{1}|^{2}+|g_{2}|^{2}e^{-2i\varphi_{0}}+g_{1}g_{2}^{*}e^{-i\varphi_{0}}\\
&+g_{2}g_{1}^{*}e^{-i\varphi_{0}}\Big]\Big(g_{1}^{*}+g_{2}^{*}e^{i\varphi_{0}}\Big)\Big\},
\end{aligned}
\end{equation}
while the denominator can be written as
\begin{equation}
\begin{aligned}
&(g^*_1+g^*_2e^{-i\phi_0})
\Big\{ (k-\Delta)- 2\pi i\big[|g_{1}|^{2}+|g_{2}|^{2}+ g_{1}g_{2}^{*}e^{-i\varphi_{0}}\\
&+ g_{1}^{*}g_{2}e^{-i\varphi_{0}} \big] \Big\}
\Big\{ (k-\Delta)+ 2\pi i\big[|g_{1}|^{2}+|g_{2}|^{2}+ g_{1}g_{2}^{*}e^{i\varphi_{0}}
\\&+ g_{1}^{*}g_{2}e^{i\varphi_{0}} \big] \Big\}.
\end{aligned}
\end{equation}
The product of the first two terms is identical to the expression inside the braces in the numerator. Thus, we have 
\begin{equation}
\frac{ g_{1}^{*} + g^*_{2} e^{-i\varphi_{0}} }{ (k-\Delta)+ 2\pi i\big[|g_{1}|^{2}+|g_{2}|^{2}+ g_{1}g_{2}^{*}e^{i\varphi_{0}}+ g_{1}^{*}g_{2}e^{i\varphi_{0}} \big] },
\end{equation}
which is the right side of Eq.~\eqref{eq:tsrs to s}.

Combining Eq.~\eqref{eq:s to t} and Eq.~\eqref{eq:tsrs to s}, Eq.~\eqref{eq:rrrr unsimplified} can be simplified into 
\begin{equation}
\begin{aligned}
&\langle0|r_{\mathrm{out}}(p_1)r_{\mathrm{out}}(p_2)r_{\mathrm{in}}^{\dagger}(k_1)r_{\mathrm{in}}^{\dagger}(k_2)|0\rangle\\
=&t_r({k_2})
t_r({k_{1}})\delta(k_{2}-p_{1})\delta(k_{1}-p_{2})\\
&+t_r({k_{1}})t_r({k_{2}})\delta(k_{1}-p_{1})\delta(k_{2}-p_{2})\\
&+i\sqrt{2\pi}\big(g_2^{*}e^{-i\phi_0}+g_1^{*}\big)\frac{1}{\pi}\delta(k_{1}+k_{2}-p_{1}-p_{2})\\
&\times s_r({p_2})s_r({p_1})[s_r({k_1})+s_r({k_2})]\frac{g^*_1+g^*_2e^{-i\phi_0}}{g_1+g_2e^{i\phi_0}}.
\end{aligned}
\end{equation}
which is Eq.~\eqref{eq:rrrr-result} in the main text. 


\subsection{Derivation of Other Two-photon Scattering Matrices}
\label{sec:Derivation of Other Two-photon Scattering Matrices}

We derive the following scattering matrices in this subsection, which are
\begin{equation}
\label{eq:TSM-two}
\begin{aligned}
\langle0|\ell_{\mathrm{out}}(p_1)\ell_{\mathrm{out}}(p_2)r_{\mathrm{in}}^{\dagger}(k_1)r_{\mathrm{in}}^{\dagger}(k_2)|0\rangle,\\
\langle0|r_{\mathrm{out}}(p_1)\ell_{\mathrm{out}}(p_2)r_{\mathrm{in}}^{\dagger}(k_1)r_{\mathrm{in}}^{\dagger}(k_2)|0\rangle.
\end{aligned}
\end{equation}
These two scattering matrices correspond to both photons reflected and one photon transmitted, while another is reflected, respectively. 

The derivations follow the same strategy as that of Eq.~\eqref{eq:rrrr-result}. We again start by inserting the normalization condition Eq.~\eqref{eq:normalization condition} into the two-photon scattering matrices, which are 
\begin{equation}
\begin{aligned}
&\langle0|\ell_{\mathrm{out}}(p_1)\ell_{\mathrm{out}}(p_2)r_{\mathrm{in}}^{\dagger}(k_1)r_{\mathrm{in}}^{\dagger}(k_2)|0\rangle
=\int_{-\infty}^{\infty} \! dk\\
&\;\Big[\langle0|\ell_{\mathrm{out}}(p_1)r_{\mathrm{in}}^{\dagger}(k)|0\rangle\langle 0|r_{\mathrm{in}}(k)\ell_{\mathrm{out}}(p_2)r_{\mathrm{in}}^{\dagger}(k_1)r_{\mathrm{in}}^{\dagger}(k_2)|0\rangle\\
&+\langle0|\ell_{\mathrm{out}}(p_1)\ell_{\mathrm{in}}^{\dagger}(k)|0\rangle\langle 0|\ell_{\mathrm{in}}(k)\ell_{\mathrm{out}}(p_2)r_{\mathrm{in}}^{\dagger}(k_1)r_{\mathrm{in}}^{\dagger}(k_2)|0\rangle\Big],
\end{aligned}
\end{equation}
and
\begin{equation}
\begin{aligned}
&\langle0|r_{\mathrm{out}}(p_1)\ell_{\mathrm{out}}(p_2)r_{\mathrm{in}}^{\dagger}(k_1)r_{\mathrm{in}}^{\dagger}(k_2)|0\rangle
=\int_{-\infty}^{\infty} \! dk\\
&\;\Big[\langle0|r_{\mathrm{out}}(p_1)r_{\mathrm{in}}^{\dagger}(k)|0\rangle\langle 0|r_{\mathrm{in}}(k)\ell_{\mathrm{out}}(p_2)r_{\mathrm{in}}^{\dagger}(k_1)r_{\mathrm{in}}^{\dagger}(k_2)|0\rangle\\
&+\langle0|r_{\mathrm{out}}(p_1)\ell_{\mathrm{in}}^{\dagger}(k)|0\rangle\langle 0|\ell_{\mathrm{in}}(k)\ell_{\mathrm{out}}(p_2)r_{\mathrm{in}}^{\dagger}(k_1)r_{\mathrm{in}}^{\dagger}(k_2)|0\rangle\Big].
\end{aligned}
\end{equation}
By replacing the single-photon scattering matrices and using the input–output relations after Fourier transformation, we obtain
\begin{equation}
\begin{aligned}
&\langle 0|\ell_{\mathrm{out}}(p_1)\ell_{\mathrm{out}}(p_2)r_{\mathrm{in}}^{\dagger}(k_1)r_{\mathrm{in}}^{\dagger}(k_2)|0\rangle
=\frac{1}{\sqrt{2\pi}}\int dt e^{ip_2t}\\
&\Big[ r_r(p_1)\big\langle 0\big| r_{\mathrm{in}}(p_1)\ell_{\mathrm{in}}(t)r_{\mathrm{in}}^{\dagger}(k_1)r_{\mathrm{in}}^{\dagger}(k_2)\big|0\big\rangle-r_r(p_1)\\
&\times i\sqrt{2\pi}\big(g_2^{*}e^{i\phi_0}+g_1^{*}\big)\big\langle 0\big| r_{\mathrm{in}}(p_1)\sigma_{-}(t)r_{\mathrm{in}}^{\dagger}(k_1)r_{\mathrm{in}}^{\dagger}(k_2)\big|0\big\rangle\\
&+ t_{\ell}(p_1)\big\langle 0\big| \ell_{\mathrm{in}}(p_1)\ell_{\mathrm{in}}(t)r_{\mathrm{in}}^{\dagger}(k_1)r_{\mathrm{in}}^{\dagger}(k_2)\big|0\big\rangle- t_{\ell}(p_1)\\
&\times i\sqrt{2\pi}\big(g_2^{*}e^{i\phi_0}+g_1^{*}\big)\big\langle 0\big| \ell_{\mathrm{in}}(p_1)\sigma_{-}(t)r_{\mathrm{in}}^{\dagger}(k_1)r_{\mathrm{in}}^{\dagger}(k_2)\big|0\big\rangle \Big],
\end{aligned}
\end{equation}
and
\begin{equation}
\label{eq:rlrr}
\begin{aligned}
&\langle 0|r_{\mathrm{out}}(p_1)\ell_{\mathrm{out}}(p_2)r_{\mathrm{in}}^{\dagger}(k_1)r_{\mathrm{in}}^{\dagger}(k_2)|0\rangle
=\frac{1}{\sqrt{2\pi}}\int dt e^{ip_2t}\\
&\Big[ t_r(p_1)\big\langle 0\big| r_{\mathrm{in}}(p_1)\ell_{\mathrm{in}}(t)r_{\mathrm{in}}^{\dagger}(k_1)r_{\mathrm{in}}^{\dagger}(k_2)\big|0\big\rangle-t_r(p_1)\\
& \times i\sqrt{2\pi}\big(g_2^{*}e^{i\phi_0}+g_1^{*}\big)\big\langle 0\big| r_{\mathrm{in}}(p_1)\sigma_{-}(t)r_{\mathrm{in}}^{\dagger}(k_1)r_{\mathrm{in}}^{\dagger}(k_2)\big|0\big\rangle\\
&+ r_{\ell}(p_1)\big\langle 0\big| \ell_{\mathrm{in}}(p_1)\ell_{\mathrm{in}}(t)r_{\mathrm{in}}^{\dagger}(k_1)r_{\mathrm{in}}^{\dagger}(k_2)\big|0\big\rangle- r_{\ell}(p_1)\\
& \times i\sqrt{2\pi}\big(g_2^{*}e^{i\phi_0}+g_1^{*}\big)\big\langle 0\big| \ell_{\mathrm{in}}(p_1)\sigma_{-}(t)r_{\mathrm{in}}^{\dagger}(k_1)r_{\mathrm{in}}^{\dagger}(k_2)\big|0\big\rangle \Big].
\end{aligned}
\end{equation}
While
\begin{equation}
\begin{aligned}
&\big\langle 0\big| r_{\mathrm{in}}(p_1)\ell_{\mathrm{in}}(t)r_{\mathrm{in}}^{\dagger}(k_1)r_{\mathrm{in}}^{\dagger}(k_2)\big|0\big\rangle=0,\\
&\big\langle 0\big| \ell_{\mathrm{in}}(p_1)r_{\mathrm{in}}(t)r_{\mathrm{in}}^{\dagger}(k_1)r_{\mathrm{in}}^{\dagger}(k_2)\big|0\big\rangle=0,\\
&\big\langle 0\big| \ell_{\mathrm{in}}(p_1)\ell_{\mathrm{in}}(t)r_{\mathrm{in}}^{\dagger}(k_1)r_{\mathrm{in}}^{\dagger}(k_2)\big|0\big\rangle=0,
\end{aligned}
\end{equation}
which are direct results of commutation relations. The terms that involve $\sigma_-$ are given by Eq.~\eqref{eq:rsrr} and Eq.~\eqref{eq:lsrr}. 

Combining the results above, the two-photon scattering matrices can be written into
\begin{equation}
\begin{aligned}
&\langle0|\ell_{\mathrm{out}}(p_1)\ell_{\mathrm{out}}(p_2)r_{\mathrm{in}}^{\dagger}(k_1)r_{\mathrm{in}}^{\dagger}(k_2)|0\rangle\\
=&-r_r({p_1})i\sqrt{2\pi}\big(g_2^{*}e^{i\phi_0}+g_1^{*}\big)\{-\frac{1}{\pi}\delta(k_{1}+k_{2}-p_{1}-p_{2})\\
&\times 
s_r({p_{2}})s^{*}_{r}({p_{1}})[s_r({k_1})+s_r({k_2})]\\
&+s_r({k_1})\delta(k_{2}-p_{1})\delta(k_{1}-p_{2})\\
&+s_r({k_2})\delta(k_{1}-p_{1})\delta(k_{2}-p_{2})\}\\
&-t_{\ell}(p_1)i\sqrt{2\pi}\big(g_2^{*}e^{i\phi_0}+g_1^{*}\big)\{-\frac{1}{\pi}\delta(k_{1}+k_{2}-p_{1}-p_{2})\\
&\times s_r({p_{2}})s^{*}_{\ell}({p_{1}})[s_r({k_1})+s_r({k_2})]\},
\end{aligned}
\end{equation}
and
\begin{equation}
\label{eq:rlrr unsimplified}
\begin{aligned}
&\langle0|r_{\mathrm{out}}(p_1)\ell_{\mathrm{out}}(p_2)r_{\mathrm{in}}^{\dagger}(k_1)r_{\mathrm{in}}^{\dagger}(k_2)|0\rangle\\
=&-t_r({p_1})i\sqrt{2\pi}\big(g_2^{*}e^{i\phi_0}+g_1^{*}\big)\{-\frac{1}{\pi}\delta(k_{1}+k_{2}-p_{1}-p_{2})\\
&\times s_r({p_{2}})s^{*}_{r}({p_{1}})[s_r({k_1})+s_r({k_2})]\\
&+ s_r({k_1})\delta(k_{2}-p_{1})\delta(k_{1}-p_{2})\\
&+s_r({k_2})\delta(k_{1}-p_{1})\delta(k_{2}-p_{2})\}\\
&-r_{\ell}(p_1)i\sqrt{2\pi}\big(g_2^{*}e^{i\phi_0}+g_1^{*}\big)
\{-\frac{1}{\pi}\delta(k_{1}+k_{2}-p_{1}-p_{2})\\
&\times s_r({p_{2}})s^{*}_{\ell}({p_{1}})
[s_r({k_1})+s_r({k_2})]\}.
\end{aligned}
\end{equation}
Single-photon process terms can be simplified by
\begin{equation}
\label{eq:s to r}
r_r(k)=-\sqrt{2\pi} i(g_1^*+g_2^*e^{i\phi_0})s_r(k),
\end{equation}
and Eq.~\eqref{eq:s to t}. Equation~\eqref{eq:s to r} is the reflection version of Eq.~\eqref{eq:s to t}.

The photon-photon interaction term can be simplified by
\begin{equation}
\begin{aligned}
&r_r({p_1})s_r^{*}({p_1})+t_{\ell}{(p_1)}s_r^{*}({p_1})\frac{g^*_1+g^*_2e^{i\phi_0}}{g^*_1+g^*_2e^{-i\phi_0}}\\
=&s_r({p_1})\frac{g^*_1+g^*_2e^{i\phi_0}}{g_1+g_2e^{i\phi_0}},\\
&t_r({p_1})s_r^{*}({p_1})+r_{\ell}{(p_1)}s_r^{*}({p_1})\frac{g^*_1+g^*_2e^{i\phi_0}}{g^*_1+g^*_2e^{-i\phi_0}}\\
=&s_r({p_1})\frac{g^*_1+g^*_2e^{-i\phi_0}}{g_1+g_2e^{i\phi_0}}.
\end{aligned}
\end{equation}
The proof is similar to that of Eq.~\eqref{eq:tsrs to s}, which requires some labor. 

Finally, the simplified two-photon scattering matrices are obtained, given by
\begin{equation}
\label{eq:llrr-result}
\begin{aligned}
&\langle0|\ell_{\mathrm{out}}(p_1)\ell_{\mathrm{out}}(p_2)r_{\mathrm{in}}^{\dagger}(k_1)r_{\mathrm{in}}^{\dagger}(k_2)|0\rangle\\
=&r_r({k_2})
r_r({k_{1}})\delta(k_{2}-p_{1})\delta(k_{1}-p_{2})\\
&+r_r({k_1})r_r({k_2})\delta(k_{1}-p_{1})\delta(k_{2}-p_{2})\\
&+i\sqrt{2\pi}\big(g_2^{*}e^{i\phi_0}+g_1^{*}\big)\frac{1}{\pi}\delta(k_{1}+k_{2}-p_{1}-p_{2})\\
&\times s_r({p_2})s_r({p_1})[s_r({k_1})+s_r({k_2})]\frac{g^*_1+g^*_2e^{i\phi_0}}{g_1+g_2e^{i\phi_0}},
\end{aligned}
\end{equation}
and
\begin{equation}
\label{eq:rlrr-result}
\begin{aligned}
&\langle0|r_{\mathrm{out}}(p_1)\ell_{\mathrm{out}}(p_2)r_{\mathrm{in}}^{\dagger}(k_1)r_{\mathrm{in}}^{\dagger}(k_2)|0\rangle\\
=&t_r({k_2})
r_r({k_1})\delta(k_{2}-p_{1})\delta(k_{1}-p_{2})\\
&+t_r({k_1})r_r({k_2})\delta(k_{1}-p_{1})\delta(k_{2}-p_{2})\\
&+i\sqrt{2\pi}\big(g_2^{*}e^{i\phi_0}+g_1^{*}\big)\frac{1}{\pi}\delta(k_{1}+k_{2}-p_{1}-p_{2})\\
&\times s_r({p_2})s_r({p_1})[s_r({k_1})+s_r({k_2})]\frac{g^*_1+g^*_2e^{-i\phi_0}}{g_1+g_2e^{i\phi_0}}.
\end{aligned}
\end{equation}


\section{Derivation of the Two-photon Nonlinearity }
\label{sec:Derivation of the Two-photon Nonlinearity}

In this section, we show how to derive the two-photon nonlinearity Eq.~\eqref{eq:two-photon nonlinearity}. We start from Eq.~\eqref{eq:output result1}. After substitution, the unsimplified result is
\begin{equation}
\label{eq:output result2}
\begin{aligned}
\psi_{\mu_1}^{(1)}(t)=&\frac{1}{\sqrt{2\pi}}\int dk e^{-ikt}f(k)\chi^{\mu_1}(k),\\
\psi_{\mu_2\mu_2^{\prime}}^{(2)}(t,t+\tau)
=&\frac{1}{2\pi}\int\int\frac{dp_{1}dp_{2}}{\sqrt{2}}
\int\int dk_{1}dk_{2}\\
&\Big\{e^{-i[p_1t+p_2(t+\tau)]}f(k_{1})f(k_{2})\times \\
&\Big[\chi^{\mu_2}(k_1)\chi^{\mu_2'}(k_2)\delta(p_{1}-k_{1})\delta(p_{2}-k_{2})\\
&+\chi^{\mu_2}(k_2)\chi^{\mu_2'}(k_1)\delta(p_{1}-k_{2})\delta(p_{2}-k_{1})\\
&+B_{\mu_2\mu_2^{\prime}}\delta(p_{1}+p_{2}-k_{1}-k_{2})
\Big]\Big\}.
\end{aligned}
\end{equation}
The $B_{\mu_2\mu_2^{\prime}}$ is used to denote the coefficient of the photon–photon interaction term in different two-photon scattering matrices. The detailed expressions of $B_{\mu_2\mu_2^{\prime}}$ are 
\begin{equation}
\label{eq:B all}
\begin{aligned}
B_{r\ell}=&i\sqrt{2\pi}\big(g_2^{*}e^{i\phi_0}+g_1^{*}\big)\frac{1}{\pi}\\
&\times s_r({p_2})s_r({p_1})[s_r({k_1})+s_r({k_2})]\frac{g^*_1+g^*_2e^{-i\phi_0}}{g_1+g_2e^{i\phi_0}},\\
B_{\ell\ell}=&i\sqrt{2\pi}\big(g_2^{*}e^{i\phi_0}+g_1^{*}\big)\frac{1}{\pi}\\
&\times s_r({p_2})s_r({p_1})[s_r({k_1})+s_r({k_2})]\frac{g^*_1+g^*_2e^{i\phi_0}}{g_1+g_2e^{i\phi_0}},\\
B_{rr}=&i\sqrt{2\pi}\big(g_2^{*}e^{-i\phi_0}+g_1^{*}\big)\frac{1}{\pi}\\
&\times
s_r({p_2})s_r({p_1})[s_r({k_1})+s_r({k_2})]\frac{g^*_1+g^*_2e^{-i\phi_0}}{g_1+g_2e^{i\phi_0}}.
\end{aligned}
\end{equation}

Transforming $s_r$ into $r_r$ in Eq.~\eqref{eq:B all} with Eq.~(\ref{eq:s to r}), resulting in 
\begin{equation}
\begin{aligned}
B_{r\ell}=&\frac{1}{2\pi^2}r_r({p_2})r_r({p_1})[r_r({k_1})+r_r({k_2})]\\
&\times\frac{1}{(g_2^{*}e^{i\phi_0}+g_1^{*})^2}\frac{g^*_1+g^*_2e^{-i\phi_0}}{g_1+g_2e^{i\phi_0}},\\
B_{\ell\ell}=&\frac{1}{2\pi^2}r_r({p_2})r_r({p_1})[r_r({k_1})+r_r({k_2})]\\
&\times\frac{1}{(g_2^{*}e^{i\phi_0}+g_1^{*})^2}\frac{g^*_1+g^*_2e^{i\phi_0}}{g_1+g_2e^{i\phi_0}},\\
B_{rr}=&\frac{1}{2\pi^2}r_r({p_2})r_r({p_1})[r_r({k_1})+r_r({k_2})]\\
&\times\frac{1}{(g_2^{*}e^{i\phi_0}+g_1^{*})^3}\frac{(g^*_1+g^*_2e^{-i\phi_0})^2}{g_1+g_2e^{i\phi_0}}.
\end{aligned}
\end{equation}
Notice that for two-photon scattering matrices in an arbitrary output direction, $B_{\mu_2\mu_2^{\prime}}$ is always proportional to $r_r({p_2})r_r({p_1})[r_r({k_1})+r_r({k_2})]$. The nonlinear term in Eq.~\eqref{eq:output result2} is 
\begin{equation}
\label{eq:nonlinear term}
\begin{aligned}
N_{\mu_2\mu_2^{\prime}}(t,t+\tau)=&\frac{1}{2\pi}\int\int\frac{dp_{1}dp_{2}}{\sqrt{2}}
\int\int dk_{1}dk_{2}\\
\Big\{e^{-i[p_1t+p_2(t+\tau)]}&f(k_{1})f(k_{2})B_{\mu_2\mu_2^{\prime}}\delta(p_{1}+p_{2}-k_{1}-k_{2})
\Big\},
\end{aligned}
\end{equation}

Rewriting the two-photon nonlinearity Eq.~\eqref{eq:nonlinear term} into a more convenient form for calculation, which is 
\begin{equation}
\label{eq:nonlinear term r}
\begin{aligned}
N_{\mu_2\mu_2^{\prime}}(t,t+\tau)=&\frac{1}{2\pi}\int\int\frac{dp_{1}dp_{2}}{\sqrt{2}}
\int\int dk_{1}dk_{2}\\ 
\Big\{e^{-i[p_1t+p_2(t+\tau)]}&
r_r({p_2})r_r({p_1})[r_r({k_1})+r_r({k_2})]\\
f(k_{1})f(k_{2})&B_{c,\mu_2\mu_2^{\prime}}\delta(p_{1}+p_{2}-k_{1}-k_{2})  
\Big\},
\end{aligned}
\end{equation}
where
\begin{equation}
B_{\mu_2\mu_2^{\prime}}=B_{c,\mu_2\mu_2^{\prime}}r_r({p_2})r_r({p_1})[r_r({k_1})+r_r({k_2})].
\end{equation}
By factoring out the frequency-independent part and denoting it by $B_{c,\mu_2\mu_2^{\prime}}$, we can focus on simplifying the integral itself. We first introduce the standard parametrization process of the two outgoing frequencies $(p_{1},p_{2})$ by their mean frequency and frequency difference,
\begin{equation}
\omega\equiv\frac{p_{1}+p_{2}}{2},\qquad
\Delta_p\equiv\frac{p_{1}-p_{2}}{2},
\label{eq:A2_change_vars}
\end{equation}
hence $p_{1}=\omega+\Delta_p$ and $p_{2}=\omega-\Delta_p$. The Jacobian coefficient of this transformation gives $dp_{1}d
p_{2}=2d\omega d\Delta_p$. The nonlinear term Eq.~\eqref{eq:nonlinear term r} can be transformed into 
\begin{equation}
\label{eq:nonlinear term r delta}
\begin{aligned}
N_{\mu_2\mu_2^{\prime}}(t,t+\tau)=&\frac{1}{2\pi}\int\frac{d\Delta_p}{\sqrt{2}}
\int\int dk_{1}dk_{2}\\ 
\Big\{e^{-i[2\omega t+(\omega-\Delta_p)\tau)]}&
r_r(\omega-\Delta_p)r_r(\omega+\Delta_p)[r_r({k_1})+r_r({k_2})]\\
f(k_{1})f(k_{2})&B_{c,\mu_2\mu_2^{\prime}} 
\Big\},
\end{aligned}
\end{equation}
where $\omega=\frac{k_{1}+k_{2}}{2}$, which follows from integrating over the delta function. Factoring out terms involving $\Delta_p$ and performing the integration, we have
\begin{equation}
\label{eq:int Delta}
\begin{aligned}
&\int d\Delta_p e^{i\Delta_p\tau}r_r(\omega-\Delta_p)r_r(\omega+\Delta_p)\\
=&-\pi i Dr_r(\omega)e^{i(\omega+C)|{\tau}|},
\end{aligned}
\end{equation}
where we introduce the shorthand $r_r(\omega)=\frac{D}{\omega+C}$ to simplify the notation. Since $r_r(\omega)$ satisfies the following property 
\begin{equation}
r_r(\frac{k_1+k_2}{2})=\frac{2 r_r({k_{1}})r_r({k_{2}})}{r_r({k_{1}})+r_r({k_{2}})},
\end{equation}
and combining with Eq.~\eqref{eq:int Delta}, the nonlinear term Eq.~\eqref{eq:nonlinear term r delta} can be further simplified into 
\begin{equation}
\label{eq:nonlinear term result}
\begin{aligned}
N_{\mu_2\mu_2^{\prime}}(t,t+\tau)=&-\frac{iDe^{iC|\tau|}B_{c,\mu_2\mu_2^{\prime}} }{\sqrt{2}}
\int\int dk_{1}dk_{2}\\ 
\Big\{e^{-i\omega (2t+\tau-|\tau|)}&r_r({k_1})r_r({k_2})f(k_{1})f(k_{2})\Big\},
\end{aligned}
\end{equation}
which is Eq.~\eqref{eq:two-photon nonlinearity} in the main text after rewriting the two integrals as a squared form. For ease of reference and the subsequent analysis, we list all constants appearing in Eq.~\eqref{eq:nonlinear term result} as follows, 
\begin{equation}
\label{eq:constants}
\begin{aligned}
\omega&=\frac{k_{1}+k_{2}}{2},\\
B_{c,r\ell}&=\frac{1}{2\pi^2}(g_2^{*}e^{i\phi_0}+g_1^{*})^{-2}
\frac{g^*_1+g^*_2e^{-i\phi_0}}{g_1+g_2e^{i\phi_0}},\\
B_{c,\ell\ell}&=\frac{1}{2\pi^2}(g_2^{*}e^{i\phi_0}+g_1^{*}\big)^{-2}\frac{g^*_1+g^*_2e^{i\phi_0}}{g_1+g_2e^{i\phi_0}},\\
B_{c,rr}&=\frac{1}{2\pi^2}\big(g_2^{*}e^{-i\phi_0}+g_1^{*}\big)
[g_2^{*}e^{i\phi_0}+g_1^{*}]^{-3}
\frac{g^*_1+g^*_2e^{-i\phi_0}}{g_1+g_2e^{i\phi_0}},\\
D&=-2\pi i[|g_1|^2+|g_2|^2e^{2i\phi_0}+2\mathrm{Re}(g_1 g_2^{*})e^{i\phi_0}],\\
C&=-\Delta+2\pi i(|g_1|^2+|g_2|^2\big)+4\pi i\mathrm{Re}(g_1 g_2^{*})e^{i\phi_{0}}.
\end{aligned}
\end{equation}


\section{Supplementary Figures of Dynamic Analysis}
\label{sec:Supplementary Figures of Dynamic Analysis}

\begin{figure*}[t]
  \centering
  \includegraphics[width=\textwidth]{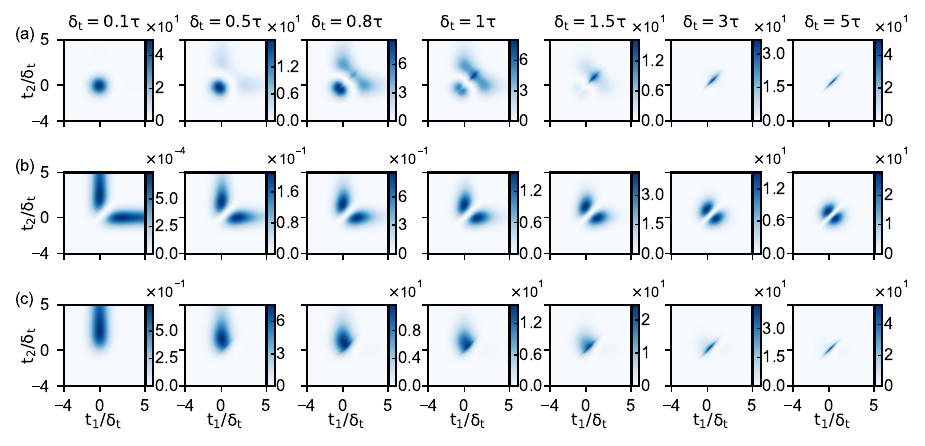}
  \caption{Unnormalized second-order correlation function for different giant atom lifetime $\tau$ relative to the pulse width $\delta_t$. (a) Both photons are transmitted. (b) Both photons are reflected. (c) One photon is transmitted, while another is reflected. The units is $|\rm{Mhz}|^2$.}
  \label{fig:fig_R1_g2_3×7_resonance_coupling}
\end{figure*}
\begin{figure*}[t]
  \centering
  \includegraphics[width=\textwidth]{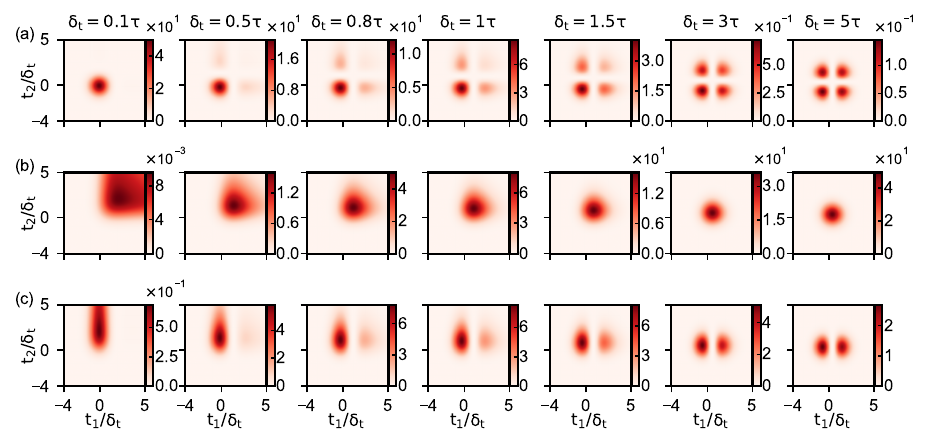}
  \caption{The product of the single-photon intensity spectrum for different giant atom lifetime $\tau$ relative to the pulse width $\delta_t$. (a) Both photons are transmitted. (b) Both photons are reflected. (c) One photon is transmitted while another is reflected. The units is $\rm{Mhz}^2$.}
  \label{fig:fig_R1_g1_3×7_resonance_coupling}
\end{figure*}

In this section, we provide results that involve more ratios for three output cases as a supplementary illustration of Fig.~\ref{fig:fig_R1_g1g2_tt_resonance_coupling}. The results are illustrated in Fig.~\ref{fig:fig_R1_g2_3×7_resonance_coupling} and Fig.~\ref{fig:fig_R1_g1_3×7_resonance_coupling}, which present a more complete picture of how the product of the single-photon intensity spectrum and the unnormalized second-order correlation function evolves as $\tau$ decreases. We find that the reflection spectrum in Fig.~\ref{fig:fig_R1_g1_3×7_resonance_coupling} (b) remains a single-photon wavepacket. The reason is that the giant atom can emit only one photon at a time. A photon can appear in the reflected channel only after undergoing an absorption–emission process. Therefore, even though $t_1$ and $t_2$ vary, the equal-time second-order correlation function in reflection still exhibits antibunching. 

Additionally, as the giant-atom lifetime $\tau$ becomes shorter (e.g., for $\delta_t = 3\tau$ and $\delta_t = 5\tau$), the outputs in all three cases gradually approach a steady form. This behavior arises because the bound state contribution becomes dominant in this regime, as analyzed in Sec.~\ref{sec:section3-2}. 

\end{document}